\documentclass{aa}  

\usepackage{graphicx}
\usepackage{txfonts}
\usepackage{booktabs}
\usepackage{titlesec}
\usepackage[dvipsnames]{xcolor}
\usepackage{subfigure}
\usepackage{pdflscape}
\usepackage{ulem}

\usepackage{color}
\usepackage{natbib,twoopt}
\usepackage[hyphenbreaks]{breakurl}
\usepackage[breaklinks]{hyperref}      
\bibpunct{(}{)}{;}{a}{}{,}             
\definecolor{cobalt}{rgb}{0.06, 0.2, 0.65}
\hypersetup{
  colorlinks,
  citecolor=blue,
  linkcolor=blue,
  urlcolor=blue,
}
\makeatletter
  \newcommandtwoopt{\citeads}[3][][]{\href{http://adsabs.harvard.edu/abs/#3}%
    {\def\hyper@linkstart##1##2{}%
     \let\hyper@linkend\@empty\citealp[#1][#2]{#3}}}
  \newcommandtwoopt{\citepads}[3][][]{\href{http://adsabs.harvard.edu/abs/#3}%
    {\def\hyper@linkstart##1##2{}%
     \let\hyper@linkend\@empty\citep[#1][#2]{#3}}}
  \newcommandtwoopt{\citetads}[3][][]{\href{http://adsabs.harvard.edu/abs/#3}%
    {\def\hyper@linkstart##1##2{}%
     \let\hyper@linkend\@empty\citet[#1][#2]{#3}}}
  \newcommandtwoopt{\citeyearads}[3][][]%
    {\href{http://adsabs.harvard.edu/abs/#3}
    {\def\hyper@linkstart##1##2{}%
     \let\hyper@linkend\@empty\citeyear[#1][#2]{#3}}}
\makeatother

\titleformat{\paragraph}
{\normalfont\normalsize}{\theparagraph}{1em}{}
\titlespacing*{\paragraph}
{0pt}{3.25ex plus 1ex minus .2ex}{1.5ex plus .2ex}

\newcommand{\bse}{\textsc{bse}\xspace}
\newcommand{\mocca}{\textsc{mocca}\xspace}

\newcommand{\fewbody}{\textsc{fewbody}\xspace}
\newcommand{\nbody}{\textsc{nbody}\xspace}

\newcommand{\fg}{\rm{1P}\xspace}
\newcommand{\sg}{\rm{2P}\xspace}
\titlerunning{MOCCA-III: Evolution of multiple stellar populations}
\authorrunning{M. Giersz et al.}

\begin{document}

   \title{MOCCA: Effects of pristine gas accretion and cluster migration on globular cluster evolution, global parameters, and multiple stellar populations}

   \author{
    M. Giersz$^{1}$,
    A. Askar$^{1}$, A Hypki$^{1,2}$, J. Hong$^{4}$, G. Wiktorowicz$^{1}$, L. Hellström$^{1}$
    }

   \institute{Nicolaus Copernicus Astronomical Center, Polish Academy of Sciences, ul. Bartycka 18, PL-00-716 Warsaw, Poland
   \\
              \email{mig@camk.edu.pl}
         \and
 Faculty of Mathematics and Computer Science, A. Mickiewicz University, Uniwersytetu Pozna\'nskiego 4, 61-614 Pozna\'n, Poland
         \and
Korea Astronomy and Space Science Institute, Daejeon 34055, Republic of Korea         
             }

   \date{Accepted XXX. Received YYY; in original form ZZZ}

   \abstract{Using the \mocca code, we study the evolution of globular clusters (GCs) with multiple stellar populations. For this purpose, the \mocca code has been significantly extended to take into account the formation of an enriched population of stars from re-accreted gas with a time delay after the formation of the pristine population of stars. The possibility of cluster migration in the host galaxy and the fact that the pristine population can be described by a model not in virial equilibrium are also taken into account. Gas re-accretion and cluster migration have a decisive impact on the observational parameters of clusters and the ratio of the number of objects between the pristine and enriched populations. The obtained results, together with observational data, suggest a speculative refinement of the AGB scenario that makes it possible to explain some observational data, such as the ratio of the pristine to the enriched populations, the observational fact that for some GCs the pristine population is more concentrated than the enriched one, and possibly a correlation between the ratio of the number of enriched stars to the total number of stars and the mass of the cluster. In this scenario, it is important to take into account the environment in which the cluster lives, the conditions in the galaxy when it formed, and the fact that a significant part of the GCs associated with the Galaxy come from dwarf galaxies that merged with the Milky Way. The initial conditions of GCs in our simulations differ from the widely used typical models, as they require GCs to fill the Roche lobe rather than being highly concentrated within it, imposing strong constraints on their formation locations within the galaxy.}


   \maketitle

%

\section{Introduction}
\label{s:intro}

Globular clusters (GCs) have for many decades been thought of as gravitationally bound large collections of stars that formed over a short period of time and have a uniform chemical composition. Observations made in recent decades, both spectroscopic and photometric, have shown that GCs contain different stellar populations that differ significantly in chemical composition, especially in their light element content. A thorough description of the observational data and their analysis leading to the explanation of the phenomenon of multiple stellar populations (MSPs) can be found in review papers (\citep[e.g.,][]{Gratton2019, Bastian2018, Milone2022}).

Observations, both spectroscopic and photometric, provide the following information:
\begin{itemize}
\item Star-to-star light element abundance variations: He, C, O, N, Na, Mg, and Al. Abundances are correlated (Na-N and N/Na-He) or anticorrelated (Na-O and N-C). This points out to hot hydrogen burning through CNO-, NeNa-, and MgAl cycles  \citep[e.g.,][]{Langeretal1993}. The processed material from these reactions is ejected by massive stars and mixes with pristine gas, resulting in enriched gas from which the second population (\sg) of stars forms. In this context, the first population (\fg) refers to stars formed from the original, unprocessed gas, while the second population (\sg) forms from gas enriched by the products of hot hydrogen burning.
Such correlations and anticorrelations
are present for the whole range of observed stellar masses, from the bottom of the main sequence (MS) up to red giant branch (RGB) and horizontal branch (HB) \citep{Dondoglioetal2021, Marinoetal2024};
\item	There are very small Fe spreads allowed of about $\sim 0.1$ dex \citep[e.g.,][]{Marinoetal2019}. There are some exceptions, such as NGC 6341 \citep[][]{Lee2023}. Such clusters are called type II GCs;
\item	There is practically no time spread between populations. It should be at most a few hundred million years \citep[e.g.,][]{Nardielloetal2015,Lucertini2021};
\item	The enriched population (\sg) is centrally concentrated, but there are some exceptions \citep[e.g.,][]{Leitingeretal2023}. \citet{Cadelanoetal2024} challenged these results for one such cluster, NGC 3201, based on observational studies of the structural and kinematic properties of the two populations, providing support that \sg, when it formed, was more centrally concentrated. One of the possible origins of the reversed spatial distribution is a merger of clusters \citep[e.g.,][]{Lee2015};
\item Multiple stellar populations are very evident in massive (heavier than  $\sim 10^4 \mathrm{M_{\odot}}$) \citep[e.g.,][and references therein]{Bastian2018} and old clusters and are observed not only in the Milky Way (MW) but also in nearby dwarf galaxies. It seems that massive young clusters that are younger than $\sim 2$ Gyr do not contain MSPs \citep[e.g.,][]{Niederhoferetal2017, Martocchiaetal2017}. The observational ratio between the number of stars from \sg to the total number of stars in the cluster ($\mathrm{N_2/N_{tot}}$) is between 0.3--0.4 and 0.9;
\item The 2P fraction increases with the mass of the cluster and probably with its age \citep[e.g.,][]{Miloneetal2020}. Such a correlation is very difficult to explain on the basis of any MSP formation scenarios;
\item	The MSPs are also related to differences in kinematic properties and binary content between populations. \sg stars have more radially anisotropic velocity distributions  \citep[e.g.,][]{Libralatoetal2019, Libralatoetal2023}, a more rapid rotation \citep[e.g.,][]{Kamannetal2020a, Cordonietal2020, Szigetietal2021}, and a lower fraction of binaries \citep[e.g.,][]{Lucatelloetal2015, Kamannetal2020b, Miloneetal2020}.
\end{itemize}

The above observational findings impose strong constraints on models explaining the formation of MSPs in GCs. The observed correlations and anticorrelations between light elements suggest the need for dilution of the ejected products of thermonuclear reactions from stars and binary systems with remnants of the gas after pristine population (\fg) formation \citep[e.g.,][]{Bastian2018}. The observed large $\mathrm{N_2/N_{tot}}$ ratios suggest a very efficient and rapid process of \fg star loss from the cluster. It is estimated that \fg's mass loss should be about 90\% of its initial mass \citep[e.g.,][]{Bastian2018}. The observed correlation between the current GC mass and the $\mathrm{N_2/N_{tot}}$ ratio imposes the strongest constraints on the models and probably suggests a strong influence on the environment in which GCs form and live.

Several scenarios have been proposed that attempt to explain the emergence and continued evolution of MSPs. These models can be divided into two groups. The first group, in which the formation of \sg is shifted in time relative to \fg and requires the re-accretion of gas by the GC and its mixing with matter ejected by AGB stars, belongs to the AGB scenario \citep[e.g.,][]{DAntonaetal2002, DErcoleetal2008, DAntonaetal2016, Caluraetal2019}. The second group belongs to models in which \sg is formed at virtually the same time as \fg and does not require the re-accretion of gas by the GC. Matter ejected from massive stars and binary systems mixes with residual primordial gas. These scenarios include: interacting massive binaries \citep[][]{deMinketal2009}, fast-rotating massive stars \citep[][]{Prantzos2006, Decressinetal2007}, early disk accretion \citep[][]{Bastianetal2013}, very massive stars \citep[][]{Vink2018, Higginsetal2023}, very massive stars due to runaway collisions \citep[][]{Denissenkov2014, Gielesetal2018}, nucleosynthesis in accretion disks around stellar-mass black holes \citep{Breen2018,Freour2024}, and the recently renewed single-binary composite scenario \citep[][]{Vanbeverenetal2012, Bekki2023}. It has also been proposed that \fg stars moving through a medium polluted by AGB and massive star ejecta could form and subsequently accrete enriched substellar companions \citep{Winter2023}. Some of these scenarios will be discussed in detail later in Section \ref{s:Discussion}. Unfortunately, as is pointed out in the review paper by \citet{Bastian2018}, no scenario can explain a significant number of observational facts.  

In recent years, there have been many attempts and simulations with various numerical codes to obtain a model corresponding to the MSP observations. Monte Carlo \citep[e.g.,][]{Vesperinietal2021, Sollima2021, Hypkietal2022, Hypkietal2025}, \nbody \citep[e.g.,][]{Lacchinetal2024}, hydrodynamic \citep[e.g.,][]{Caluraetal2019, McKenzie2021, Yaghoobietal2022}, and semi-analytical \citep[e.g.,][]{Bekki2023, Parmentier2024} codes have been used for this purpose. Despite the success in reproducing some observational properties of GCs from MSPs, these simulations were associated with important simplifying assumptions about the effect of gas re-accretion on GCs, the time of formation, and parameters describing \sg, as well as the effect of external conditions on the evolution of GCs. Because of the above, it is important to check whether the results obtained so far concerning the mass, half-mass radius ($\mathrm{R_h}$) of GCs, or $\mathrm{N_2/N_{tot}}$ ratio will be maintained when a more physical prescription of \sg formation is introduced and the influence of the external environment is taken into account, at least partially and in a simplified manner.

To test how the inclusion of environmental effects and the effect of mass re-accretion on GCs and the time shift of \sg formation affect the evolution of stellar populations in GCs and whether the constraints and results obtained from previous simulations are still correct, the \mocca code was substantially updated. A detailed description of the code and its recent upgrades is given in Section \ref{s:Method}. The expansion of the \mocca code will allow us to follow in a simplified way the evolution of GCs with MSPs in a variable galactic environment, the influence of the parameters describing the formation of \fg and \sg, and the parameters defining their spatial and kinematic characteristics.  We expect that these extensions will allow us to better understand the evolution of GCs with MSPs, bring closer a more accurate reproduction of the observations, and impose stronger constraints on the initial conditions and physical processes responsible for the formation of MSPs in GCs. The aim of this work is to investigate how parameters related to the inclusion of MSPs influence the evolution of GCs within their galactic environment and shape present-day MSP fractions and other cluster properties.

The paper is organized as follows. Section \ref{s:Method} describes the latest version of the \mocca code with all added extensions and the initial conditions of the numerical simulations performed for this paper. Section \ref{s:Results} presents the results obtained with the \mocca simulations through the ratio $\mathrm{N_2/N_{tot}}$, $\mathrm{R_h}$, and cluster mass. In Section \ref{s:Discussion} we discuss the potential implications obtained from \mocca simulations for the observational signatures of the multiple populations in GCs and some implications for the scenarios of their formation. We also provide a speculative refinement of the AGB scenario explaining the observational parameters of MWGCs and the correlation between cluster mass and the $\mathrm{N_2/N_{tot}}$ ratio. Section \ref{s:Conclu} briefly summarizes the main findings of the paper and describes future works.

\section{Method}
\label{s:Method}
This section presents the description of the \mocca code with all additions needed to follow the evolution of GCs with re-accreted gas. All parameters needed to describe the MSPs and GCs are also discussed here. 

\subsection{MOCCA code}
\label{s:code}
This work is based on the numerical simulations performed with the \mocca Monte Carlo code \citep[][]{Giersz1998, Hypki2013, Gierszetal2013, Hypkietal2022, Hypkietal2025}, which is based on H{\'e}non's Monte Carlo method \citep[][]{Henon1971, Stodolkiewicz1982}. \mocca is an advanced code that performs full stellar and dynamical evolution of real-size star clusters up to the Hubble time. \mocca can follow the full dynamical and stellar evolution of MSPs.  For mergers and mass transfers between stars from different populations, we only assume that they form a so-called mixed population because we do not provide procedures to accurately model the chemical mixing between two stars belonging to different populations. Strong dynamical interactions in \mocca are handled using the \fewbody code \citep[][]{Fregeauetal2004, Fregeau2007}. In its current implementation within \mocca, \fewbody does not account for dissipative effects, which are physical mechanisms that lead to energy loss during encounters, such as tidal dissipation \citep[][]{PressTeukolsky1977, MardlingAarseth2001, Hellstrometal2022} or energy loss due to gravitational wave (GW) emission \citep[][]{Peters1964, Hansen1972, Samsingetal2018, Tranietal2022}.

The implementation of stellar and binary evolution within the \mocca code is based on the rapid population synthesis code \bse code \citep[][]{Hurleyetal2000, Hurleyetal2002} that has been strongly updated by \citet[][]{Bellonietal2017, Bellonietal2018}, \citet[][]{Banerjee2020}, and \citet[][]{Kamlahetal2022}. The stellar and binary evolution parameters used in the presented simulations are referred to as Model C in \citet{Kamlahetal2022}. In short, the metallicity of both populations in all the simulated models was set to $Z = 0.001$. The updated treatment for the evolution of massive stars was used according to \citet{Tanikawaetal2020} together with an improved
treatment for mass loss due to stellar winds and the inclusion of pair and pulsational pair-instability supernova \citep[][]{Belczynskietal2016}. The masses of black holes (BHs) and neutron stars (NSs) were determined according to the rapid supernovae prescriptions from \citet{Fryeretal2012}. The NS natal kicks were sampled from a Maxwelllian distribution with $\sigma = 265$ km/s \citep{Hobbsetal2005}. However, for BHs, these natal kicks were reduced according to the mass fallback prescription \citep[][]{Belczynskietal2002, Fryeretal2012}. The formation of NSs with negligible natal kicks through electron-capture or accretion-induced supernova was also enabled. Another feature of these models is the inclusion of GW recoil kicks whenever two BHs merge \citep[][]{Bakeretal2008, Morawskietal2018}). We assumed low birth spins for BHs with uniformly sampled values between 0 and 0.1 \citep[][]{Fuller2019}. The orientation of the BH spin relative to the binary orbit is randomly distributed \citep{Morawskietal2018}.

The Galactic potential was modeled in the \mocca code as a simple point-mass, taking as the central mass the value of the Galaxy mass enclosed within the GC’s circular orbit. The rotation velocity of the MW was set at \rm{220} km/s over the entire range of Galactocentric distances. This assumption allows us to calculate the cluster tidal radius ($R_t$)  as a function of the cluster Galactocentric position and its mass. The code treats the escape process in tidally limited clusters in a realistic manner as described in \citet{Fukushige2000}. Here, the escape of an object from the system is not instantaneous, but delayed in time. The assumption about point-mass Galactic potential is used as a standard option in most N-body and Monte Carlo simulations, including simulations of MSPs by \citet{Vesperinietal2021, Hypkietal2022, Hypkietal2025}. 

\subsubsection{MSPs in the MOCCA code}
\label{s:nTD}
In two previous papers \citep[][]{Hypkietal2022, Hypkietal2025}, the evolution of GCs with MSPs under the asymptotic giant branch (AGB) scenario \citep[e.g.,][]{DErcoleetal2008, DErcoleetal2010, DErcoleetal2012, Conroy2011, DAntonaetal2016, Caluraetal2019} was studied using the \mocca code \citep[][]{Hypki2013, Gierszetal2013, Kamlahetal2022}. In those simulations, it was assumed for simplicity that the process of formation of distinct populations of stars in GCs occurs simultaneously. At the start of the simulations, the two stellar populations are already formed and are together in virial equilibrium (hereinafter, the model will be referred to as no time delay MSPs - nTD-MSPs). Simulations have shown that models of clusters that are close to the MW center, in which the \fg fills the Roche lobe and the \sg is strongly concentrated toward the center reproduce very well the ranges of the ratio $\mathrm{N_2/N_{tot}}$ and the ranges of global observational parameters of MWGCs, such as the cluster mass and its $\mathrm{R_h}$. Despite these successes, the nTD-MSP model implemented in the \mocca code had very serious drawbacks; namely, it did not take into account the fact that the formation of \sg is delayed in time relative to the time of \fg formation and the re-accretion of gas surrounding the just-formed cluster, which then mixes with enriched material from the ejected envelopes of AGB stars. To address these problems, the \mocca code has been expanded with the inclusion of the following features:
\begin{enumerate}
    \item Time delay for forming \sg stars relative to \fg stars. The standard value used in the paper is \rm{100} Myr \citep[][and references therein]{Caluraetal2019}. We assume that the formation process is instantaneous and that all accreted gas is entirely converted into stars; 
    \item Time delay in re-accretion of the gas surrounding the cluster. The standard value used in the paper is \rm{50} Myr, which is about \rm{20 - 30} Myr after the residual gas is removed from the cluster due to \fg supernovae explosions and about \rm{10} Myr after AGB stars start to pollute the cluster environment \citep[][and references therein]{Caluraetal2019}; 
    \item Consideration of the mass of ejected AGB star envelopes in the formation process of \sg stars;
    \item Migration of GCs to larger Galactocentric distances. We set it according to results presented by \citet[][see Figure 1 there]{Meng2022}, where GCs migrate in the first \rm{1} Gyr of their evolution to Galactocentric distances larger than about \rm{1 - 2} kpc.
\end{enumerate}
The new model will be referred to as time-delay MSPs (TD-MSPs). 

In addition, the process of cluster migration due to dynamical friction has also been implemented in the \mocca code, based on the work of \citet{ArcaSedda2014}. In this paper, simulation results from cluster evolution with dynamical friction included are not discussed. 
Dynamical friction within the MW significantly affects cluster evolution only for massive clusters with Galactocentric distances less than \rm{2} kpc. Therefore, it does not play a significant role in the long-term evolution of the clusters discussed in this article, assuming that their migration from low to large Galactocentric radii occurs within the first gigayear \citep{Meng2022}.

Implementing the new functionality in the MOCCA code required quite deep changes related to the fact that \sg objects\footnote{please keep in mind that the term object means both stars and binary systems and will be used in this sense throughout the rest of the article} are treated as gas rather than stars for a certain time (\sg formation delay time – $\mathrm{t_{delay}}$). In the following subsections, we describe, step by step, the changes made to the \mocca code.

\subsubsection{\sg, gas re-accretion, and star formation}
\label{s:reaccretion}
To reflect the complex process of gas re-accretion onto a GC, we made the following simplifying assumptions. The parameters of \fg were determined by: the initial model (usually \citet{King1966} model), the Galactocentric distance ($\rm R_g$ equivalent to the radius of the GC circular orbit around the center of the MW), the ratio of the tidal radius $\rm (\mathrm{R_t}/pc = 0.31 (R_g/kpc) (M_g/M_{\odot})^{1/3}$, where we used constant MW rotation velocity ($\rm V_g$) equal to \rm{220} km/s) to the half-mass radius ($\mathrm{R_{h1}}$ radius containing half of the cluster mass, which is equal to the \fg mass), the initial mass function (IMF adopted as given in \citet{Kroupa2001} with the minimum mass equal to \rm{0.08} $\rm M_{\odot}$ for both populations), including the maximum stellar mass ($\mathrm{m_{1max}}$ set for \fg to \rm{150} $\rm M_{\odot}$), and the virial ratio ($\mathrm{Q_1}$ defined as the ratio between the cluster kinetic energy to the absolute value of the cluster potential energy. If it is equal to \rm{0.5} then the cluster is in so-called virial equilibrium), which can be different than \\rm{0.5}. \sg is drawn in a similar manner to \fg with one exception. \sg's concentration is controlled by a parameter $\mathrm{conc_{pop}}$ equal to the ratio of \sg's half-mass radius to \fg's half-mass radius ($\mathrm{conc_{pop} = R_{h2}/R_{h1}}$). Both populations are not forced to be together in virial equilibrium as in the case of the nTD-MSP model. \sg until time $\mathrm{t_{delay}}$ does not change its spatial distribution, the position of each star/binary system, treated as a gas particle, does not change over time, it is constant. Only \fg objects undergo dynamical and stellar evolution. \sg objects, up to the time $\mathrm{t_{delay}}$, contribute only as a source of gravitational potential and do not participate in any interactions with \fg objects.

In the \mocca code, the simulation of the complex pristine gas re-accretion process is controlled by four parameters; namely, the initial time at which gas re-accretion starts ($\mathrm{t_{start}}$), the time of completion of the accretion process ($\mathrm{t_{end}}$), which is usually equal to the $\mathrm{t_{delay}}$, the mass of the pristine gas that remained in the cluster after \fg formation ($\mathrm{M_{start}}$), and the total mass of accreted gas ($\mathrm{M_{end}}$). In addition to the mass of accreted gas, the mass of ejected matter from AGB stars is added on an ongoing basis as it is ejected from AGB stars during stellar evolution. The AGB wind has a velocity lower than 10 – 30  km/s \citep[][]{Loupetal1993, Yasudaetal2019}. So, it is assumed that ejected material will stay in the cluster and accumulate in its center. From a technical point of view, all \sg objects up to $\mathrm{t_{delay}}$ time are treated as variable mass gas particles. The mass of the gas particles varies linearly in time from $\mathrm{t_{start}}$ to  $\mathrm{t_{end}}$, from $\mathrm{M_{start}}$ to $\mathrm{M_{end}}$, which mimics the gas re-accretion process. After time $\mathrm{t_{delay}}$, all gas particles are instantly converted into stars and binary systems. The transformation of gas into POP2 stars follows the previously drawn population of stars and binary systems, which were initially assumed as gas particles. The total mass of gas converted into stars corresponds exactly to the mass of \sg, and the initial King model and the IMF determine the number and mass distribution of single and binary stars. The conversion of gas particles into stars occurs at the beginning of the next time step. From then on, \sg objects take part in stellar evolution, relaxation, and all types of dynamic interactions. 

\subsubsection{Cluster migration}
\label{s:migration}
Many previous works have indicated that the process of GC formation occurs in the early stages of galaxy formation when the galaxy mass quickly increases and galactic potential is highly variable and strongly influences the movement of clusters in the galaxy and their evolution \citep[e.g.,][]{Kruijssen2015, Lietal2017, Forbesetal2018, ElBadryetal2018, Meng2022, Sameieetal2023, vanDonkelaaretal2023, DeLuciaetal2024}. \citet{Meng2022} used cosmological simulations for MW-type galaxies to study the effects of the galaxy's strong and rapidly varying tidal field on the motion of clusters and the probability of their survival. They showed that GCs that formed very close to the galactic center (less than 1-2 kpc) over a period of about 1 Gyr increase their orbit by a factor of at least two. In the current version of the \mocca code, the orbit of a cluster can only be circular, and the mass of the galaxy within that orbit is concentrated at the center (point mass). 
Thus, the implementation of the migration of GCs' orbits in the \mocca code is made only for circular orbits whose size changes by a factor of $\mathrm{frac_{tidal}}$. This change is performed stepwise at the beginning of a new time step when the cluster lifetime exceeds the $\mathrm{t_{tidal}}$ critical value. In the new version of the code, the change in orbit size will be spread over time, and the fact that GCs' orbits are eccentric will be taken into account. For this purpose, the idea proposed by \citet{Caietal2016} will be used to compare the mass loss averaged over the cluster orbit for eccentric and circular orbits. They showed that for an eccentric orbit, it is possible to assign a circular orbit for which GC will have, on average, the same mass loss as in an eccentric orbit. This is a very rough solution, but it will enable the \mocca code to be used for the first time to study the evolution of clusters with eccentric orbits and in the variable tidal field of the galaxy. 

\subsection{Initial conditions}
\label{s:initial}
Initial conditions for \mocca simulations studying the evolution of GCs with MSPs can be divided into two groups; namely, the ones related to the global parameters of the cluster and the environment in which it moves, and the ones related to the technical parameters describing the stellar populations. The global parameters are: the number of objects in the cluster, the size of the GC circular orbit relative to the galactic center ($\mathrm{R_g}$), the King parameter describing the initial model ($\mathrm{W_{o1}}$), the virial ratio for \fg ($\mathrm{Q_1}$), the factors describing the migration of the cluster ($\mathrm{frac_{tidal}}$ and $\mathrm{t_{tidal}}$), whether or not the cluster initially fills the Roche lobe (TF or nTF), $\mathrm{R_h}$ (remember that at the beginning it is equal to $\mathrm{R_{h1}}$), the IMF, the upper mass limit for \fg ($\mathrm{m_{1max}}$), and the binary fraction along with the parameter distributions of binary systems. The technical parameters are: the number of \sg stars ($\mathrm{N_2}$), the King parameter describing the model ($\mathrm{W_{o2}}$), the concentration parameter, $\mathrm{conc_{pop}}$, the virial coefficients for \sg ($\mathrm{Q_2}$), the upper mass limit for \sg stars ($\mathrm{m_{2max}}$), and the parameters describing the re-accretion of gas: $\mathrm{t_{start}}$, $\mathrm{t_{end}}$, $\mathrm{t_{delay}}$, $\mathrm{M_{start}}$, and $\mathrm{M_{end}}$. The ranges of the above parameters are summarized in Table~\ref{t:Tab1}. The table requires a few comments to facilitate its analysis. It is important to emphasize that in the TD-MSP model, initially, the GC consists only of \fg stars. Thus, all initial global cluster parameters refer only to \fg. Once the evolution of \sg is started at time $\mathrm{t_{delay}}$, which in simulated models is always equal to $\mathrm{t_{end}}$, the cluster's global parameters refer to both populations, \fg and \sg. Please keep this fact in mind when analyzing the figures in Section \ref{s:Results}. The number of parameters that determine the models with MSPs is very large, and it would not be practical to run simulations with all combinations of these parameters. Only simulations with a representative subset of the parameters have been calculated, so that meaningful conclusions can be drawn about the dependence of observational GC parameters on their combinations. We computed about 100 \mocca models for this paper.

\begin{table*}[h!]
\centering
\caption{Global and technical parameters for simulations with MSPs}
\label{t:Tab1}
\resizebox{\textwidth}{!}{%
\begin{tabular}{|c|c|c|c|c|c|c|c|c|c|}
\hline
\multicolumn{10}{|c|}{\textbf{Global Parameters}} \\ \hline
$\mathrm{N_1}$ & $\mathrm{R_g}$ & $\mathrm{Q_1}$ & $\mathrm{W_{o1}}$ & $\mathrm{frac_{tidal}}$ & $\mathrm{t_{tidal}}$ (Gyr) & Tidal Field & $\mathrm{R_{h1}}$ & IMF & Binary Fraction \\ \hline
400,000   & 1   & 0.5 & 3 & 1 & 1000 & TF  & 1 & Kroupa (2001), & 95\% \\ 
800,000   & 2   & 0.6 & 4 & 2 & 1    & nTF & 2 & $\mathrm{m_{1max}=150 M_{\odot}}$ &   \\ 
1,600,000 & 4   & 0.7 & 5 &   &      &     &   &    &      \\ 
3,200,000 &     &     & 6 &   &      &     &   &    &      \\ \hline
\multicolumn{10}{|c|}{\textbf{Technical Parameters}} \\ \hline
$\mathrm{N_2}$ & $\mathrm{N_2/N_{tot}}$ & $\mathrm{W_{o2}}$ & $\mathrm{conc_{pop}}$ & $\mathrm{Q_2}$ & $\mathrm{m_{2max}} (\mathrm{M_{\odot}})$ & $\mathrm{t_{start}}$ (Myr) & $\mathrm{t_{end}}$ (Myr) & $\mathrm{M_{start} (M_{\odot}}$) & $\mathrm{M_{end} (M_{\odot}}$) \\ \hline
50,000  & 150,000 / 550,000   & 7 & 0.05 & 0.5 & 20  & $0.05 \times \mathrm{t_{end}}$ & 50          & 0 & Mass \\ 
100,000 & 50,000 / 850,000    &   & 0.1  &     & 8   & $0.5 \times \mathrm{t_{end}}$  & 100         &   & computed \\  
150,000 & 100,000 / 900,000   &   & 0.2  &     &     &                              & equal to    &   & from $\mathrm{N_2}$,\\ 
200,000 & 300,000 / 1,200,000 &   &      &     &     &                              & $\mathrm{t_{delay}}$ &   & IMF and\\ 
300,000 & 400,000 / 2,000,000 &   &      &     &     &                              &             &   & ejected by
\\ 
400,000 & 200,000 / 3,400,000 &   &      &     &     &                              &             &    & AGB stars\\ \hline
\end{tabular}%
}
\tablefoot{The meaning of the parameters is as follows: $\mathrm{N_1}$ and $\mathrm{N_2}$ – number of \fg and \sg objects, respectively; $\mathrm{W_{o1}}$ and $\mathrm{W_{o2}}$ – King model parameters for \fg and \sg, respectively; $\mathrm{Q_1}$ and $\mathrm{Q_2}$ – virial parameters; $\mathrm{R_g}$ – the size of the GC circular orbit; $\mathrm{R_{h1}}$ – half-mass radius of \fg; tidal field – tidally filling (TF) or underfilling (nTF); IMF – initial mass function; binary fraction – the ratio between the number of binaries and total objects; $\mathrm{t_{tidal}}$ – time when GC orbit size increases; $\mathrm{conc_{pop}}$ – the ratio of half-mass radii ($\mathrm{R_{h2}/R_{h1}}$); $\mathrm{m_{1max}}$ and $\mathrm{m_{2max}}$ – maximum stellar masses; $\mathrm{t_{start}}$ – initial time of gas re-accretion; $\mathrm{t_{end}}$ – completion time of accretion; $\mathrm{M_{start}}$ and $\mathrm{M_{end}}$ – mass of pristine and accreted gas, respectively. The GC masses presented in this work range from \rm{4.75}$\times 10^5 M_{\odot}$ to \rm{3.83}$\times 10^6 M_{\odot}$
The GC models presented in this work are designed to investigate how environmental factors influence the formation and evolution of MSPs. While they are not calibrated to reproduce the full diversity of MWGCs, we use them in Section \ref{s:Discussion} to draw insights into the role of the galactic environment in shaping GC evolution.}
\end{table*}

\section{Results}
\label{s:Results}
In this section, we analyze the impact of gas re-accretion, cluster migration, and each of the individual global and technical parameters, described in the previous section, on the evolution of the cluster mass, $\mathrm{R_h}$, and ratio, $\mathrm{N_2/N_{tot}}$.

\subsection{Gas re-accretion and migration}
\label{s:reacretion}
Before we begin to study the effect of global and technical parameters on the evolution of observable cluster parameters (henceforth when we talk about global observational parameters of GCs we mean the following parameters: cluster mass, $\mathrm{R_h}$, and ratio $\mathrm{N_2/N_{tot}}$), we would like to focus first on the effects that gas re-accretion and cluster migration bring to the observable parameters.
All simulations to date of the evolution of GCs with MSPs performed under the AGB framework have assumed for simplicity that: \sg is formed and evolves together with \fg \citep[][]{Vesperinietal2021, Hongetal2019, Hypkietal2022, Hypkietal2025}, \sg is formed with AGB ejecta and pristine gas left after \fg formation \citep[][]{Sollima2021}, and \sg is formed from re-accreted gas with a fixed \fg potential \citep[][]{Caluraetal2019}. Those simulations did not take into account the effects on the properties of \fg connected with the slow accumulation of re-accreted gas in the cluster center.

\begin{figure*}
    \centering
    \subfigure[]{\includegraphics[width=1.0\textwidth]{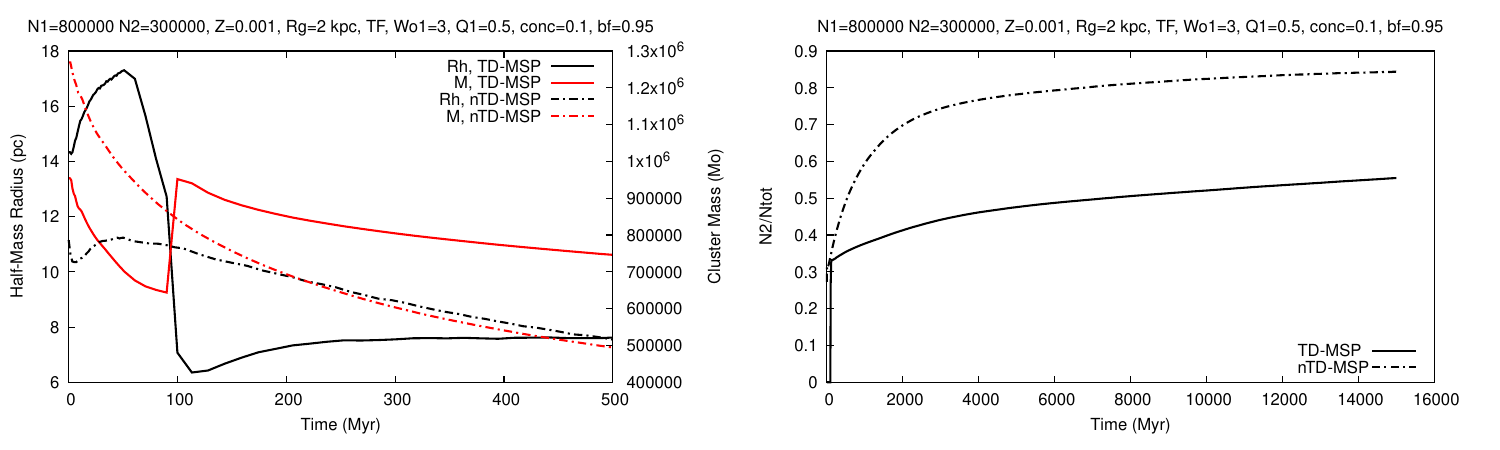}} 
    \caption{Left panel: Evolution of $\mathrm{R_h}$ for nTD-MSP model (solid red line), and TD-MSP model (solid black line), and evolution of the total cluster mass for nTD-MSP model (dash-dotted red line), and TD-MSP model (dash-dotted black line). Right panel: Evolution of the ratio $\mathrm{N_2/N_{tot}}$ for the nTD-MSP model (dash-dotted black line), and TD-MSP model (solid black line). The global cluster parameters are listed at the top of each panel: $\mathrm{N_1}$ - number of \fg objects, $\mathrm{N_2}$ - number of \sg objects, $\mathrm{Z}$ - metallicity, $\mathrm{R_g}$ - galactocentric distance (size of the circular orbit), TF - tidally filling \fg, $\mathrm{W_{o1}}$ – King parameter for \fg, $\mathrm{conc}$ - concentration parameter ($\mathrm{conc_{pop}=R_{h2}/R_{h1}}$), $\mathrm{bf}$ – binary fraction, $\mathrm{Q_1}$ - virial ratio for \fg. Unless otherwise noted, all figures from now on are made for the parameters $\mathrm{t_{start}= 0.5*t_{delay}}$ and  $\mathrm{t_{end} = t_{delay}} = 100$ Myr.}
    \label{f:Fig1} 
\end{figure*}

\begin{figure*}
    \centering
    \subfigure[]{\includegraphics[width=1.0\textwidth]{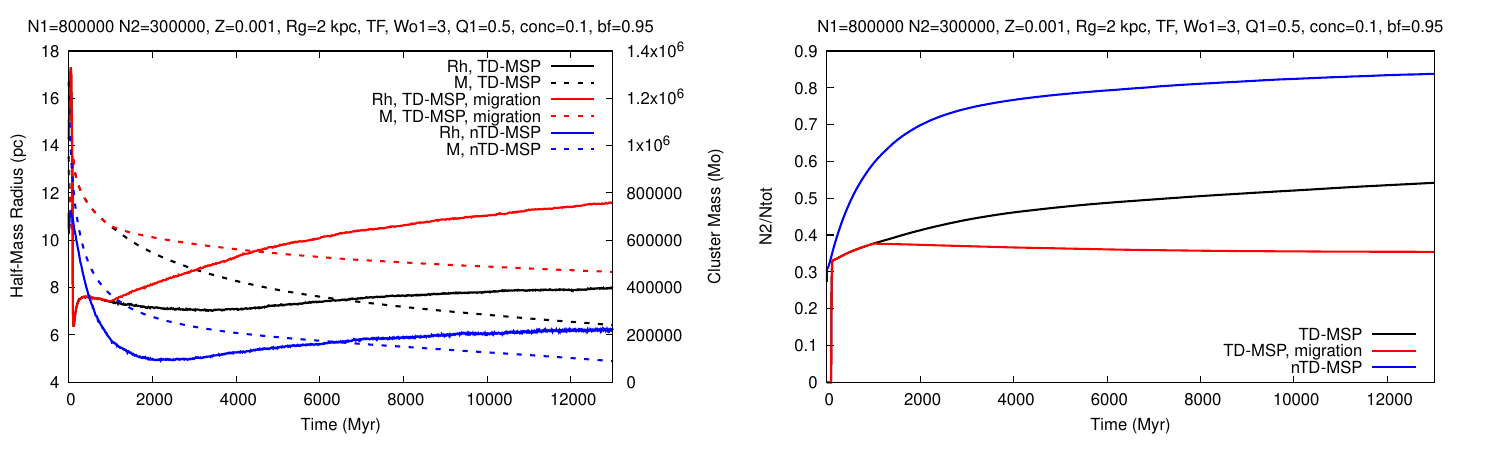}} 
    \caption{Left panel: Evolution of $\mathrm{R_h}$ for nTD-MSP model (solid blue line), TD-MSP model (solid black line), and TD-MSP model with migration (solid red line), and evolution of the total cluster mass for nTD-MSP model (dashed blue line), TD-MSP model (dashed black line), and TD-MSP model with migration (dashed red line). Right panel: Evolution of the ratio $\mathrm{N_2/N_{tot}}$ for nTD-MSP model (solid blue line), TD-MSP model (solid black line), and TD-MSP with migration model (solid red line). The global cluster parameters are listed at the top of each panel and are the same as for Figure~\ref{f:Fig1}. Note the small reduction in the value of the $\mathrm{N_2/N_{tot}}$ ratio for the model with migration. It is related to the fact that after migration, \fg objects do not escape at the same rate as before, and to the fact that mainly \sg objects are ejected from the cluster as a result of dynamic interactions.}
    \label{f:Fig2} 
\end{figure*}

The evolution of the cluster models with the gas re-accretion and cluster migration is depicted in Figure~\ref{f:Fig1} and Figure~\ref{f:Fig2}, respectively. Assuming that the GC is in quasi-dynamical equilibrium, we can, using the virial theorem, determine how the $\mathrm{R_h}$ will evolve when the cluster loses mass or gains it. When it loses mass, the characteristic radius ($\mathrm{R_h}$) should increase, and when it gains mass, it should decrease. In Figure~\ref{f:Fig1} left panel, we see exactly this behavior. First, as the cluster loses mass due to stellar evolution and tidal stripping, the $\mathrm{R_h}$ increases (solid and dashed black lines). As the re-accretion of gas begins, the $\mathrm{R_h}$ strongly decreases from 50 Myr to 100 Myr (solid black line). The decreasing $\mathrm{R_h}$ for the nTD-MSP model is associated with tidal stripping. It continues until the end of evolution and is related to the fact that the cluster is TF. For the TD-MSP model, re-accretion of gas causes the $\mathrm{R_h}$ to decrease and the cluster to become nTF. The next increase in $\mathrm{R_h}$ is related to the fact that the cluster is free to expand until it again becomes TF. The evolution of the cluster mass behaves as expected. For the nTD-MSP model, it decreases continuously due to stellar evolution and tidal stripping, For the TD-MSP model, it first decreases due to stellar evolution and tidal stripping, then increases strongly when \sg is formed at $t_{delay}$ time, and then decreases in a similar way as in the nTD-MSP model. 
We know from the previous studies \citep[e.g.,][]{Hypkietal2022, Hypkietal2025} that the ratio $\mathrm{N_2/N_{tot}}$ strongly depends on whether the cluster is TF or how strongly nTF it is.  The more nTF the cluster is, the less this ratio increases due to fewer escaping stars. So the effect of gas re-accretion should lead to much smaller observed values of this ratio. Indeed, we can observe this type of evolution in the right panel of Figure~\ref{f:Fig1}. The ratio $\mathrm{N_2/N_{tot}}$ for the model without gas re-accretion is significantly larger than for the model with gas re-accretion. This difference can be as high as 0.3.

The migration of GCs to greater galactocentric distances will be associated with an increase in the cluster's $\mathrm{R_t}$. Assuming a point mass of the galaxy and a flat rotation curve, the $\mathrm{R_t}$ is a function of galactocentric distance and cluster mass, $\mathrm{R_t \sim R_g^{2/3} (M/M_{\odot})^{1/3}}$. Increasing $\mathrm{R_t}$ will cause the cluster to no longer be TF and become nTF. The consequence will be a significant slowdown in the growth or freezing of the ratio $\mathrm{N_2/N_{tot}}$. Since the cluster is no longer limited by $\mathrm{R_t}$, it can expand freely. This will lead to an increase in $\mathrm{R_h}$ and slow the cluster's mass loss. Indeed, these predictions are confirmed in Figure~\ref{f:Fig2}. In the left panel, we can see a significant increase in $\mathrm{R_h}$ and a slowdown in mass loss when the galactocentric distance is increased. In the right panel, the ratio $\mathrm{N_2/N_{tot}}$ is practically constant after increasing the galactocentric distance. 
In conclusion, the migration of clusters to larger galactocentric distances will result in GCs having larger masses and $\mathrm{R_h}$ and a significantly smaller $\mathrm{N_2/N_{tot}}$ ratio. This is not a combination that would help explain the observed parameters of GCs. In the following sections, we try to find combinations of global and technical MSP parameters that will bring the cluster models closer to the observed parameters for MWGCs.

\subsection{Number of \fg and \sg objects}
\label{s:number}
\begin{figure*}
    \centering
    \subfigure[]{\includegraphics[width=1.0\textwidth]{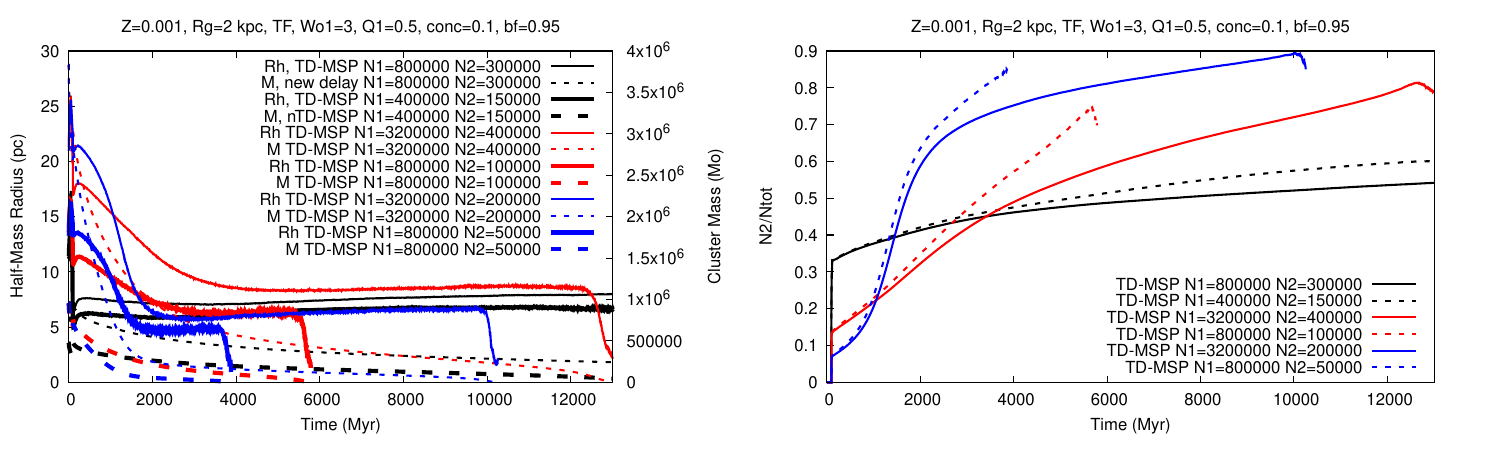}} 
    \caption{Left panel: Evolution of $\mathrm{R_h}$ (solid lines) and the cluster total mass (dashed lines) for TD-MSP models with different numbers of \fg objects and \sg objects. Right panel: Evolution of the ratio $\mathrm{N_2/N_{tot}}$ for TD-MSP models with different numbers of \fg objects and \sg objects. The global cluster parameters are listed at the top of each panel and are the same as for Figure~\ref{f:Fig1}. Note the sudden drop in $R_h$ observed in this and the following figures. It is related to the dispersion of the cluster on the dynamical timescale as a result of the presence and energy generation by the BHS \citep[][]{Gierszetal2019}.}
    \label{f:Fig3} 
\end{figure*}

Figure~\ref{f:Fig3} shows models with the different $\mathrm{N_2/N_{tot}}$ ratios for different numbers of \fg objects. Generally speaking, more massive TF clusters have a longer half-mass relaxation time, a larger $\mathrm{R_t}$ for the same galactocentric distance, and a larger $\mathrm{R_h}$. Thus, the evolution of GCs with more \fg and \sg objects will be slower, and clusters will be able to survive longer than clusters with fewer objects. Since the timescale of the evolution of clusters with fewer objects is faster, we can expect the $\mathrm{N_2/N_{tot}}$ ratio to grow faster than for clusters with more objects. The larger the number of objects, the slower the $\mathrm{N_2/N_{tot}}$ ratio grows. Indeed, the described evolution of cluster observational parameters is fully reflected in Figure~\ref{f:Fig3}. An interesting effect can be observed in the left panel of Figure~\ref{f:Fig3} for the evolution of $\mathrm{R_h}$. The smaller the initial mass of the cluster, the faster the dissolution of the cluster. The fast cluster dissolution occurs due to the presence of the black hole subsystem (BHS). The dissolution process is described in detail in the work of \citet{Gierszetal2019}. Just before dissolution, the structure of the cluster is very characteristic. The cluster has a mass of about tens of thousands $\mathrm{M_{\odot}}$ and $\mathrm{R_h}$ of a few parsecs, i.e., it is very bloated. The right panel of Figure~\ref{f:Fig3} shows the evolution of the $\mathrm{N_2/N_{tot}}$ ratio. The smaller the number of \sg objects with the same number of \fg objects, the higher the final value of the $\mathrm{N_2/N_{tot}}$ ratio. This is due to the fact that in the models presented in the paper, the King parameter for \fg is $\mathrm{W_{o1}}=3$. This means that the cluster was initially only slightly concentrated toward the center and very sensitive to tidal stripping. The small mass of \sg objects implies only a slight reduction in $\mathrm{R_h}$ due to gas reaccretion. That is when \sg is switched on, the cluster is still not highly concentrated and will lose mainly \fg objects due to tidal stripping. Note an interesting feature visible in this figure; namely, for small initial values of $\mathrm{N_2/N_{tot}}$ the increase in $\mathrm{N_2/N_{tot}}$ occurs slowly, then accelerates after about 1 Gyr of evolution to grow at the end of the cluster evolution to values characteristic of MWGCs. The small $\mathrm{N_2/N_{tot}}$ values for young clusters may be a response to observational data suggesting that young and massive star clusters (YMSC) do not exhibit \sg objects. This problem will be discussed in detail in Section~\ref{s:Discussion}.

\subsection{King parameter, $\mathrm{W_{o1}}$}
\label{s:Wo}
\begin{figure*}
    \centering
    \subfigure[]{\includegraphics[width=1.0\textwidth]{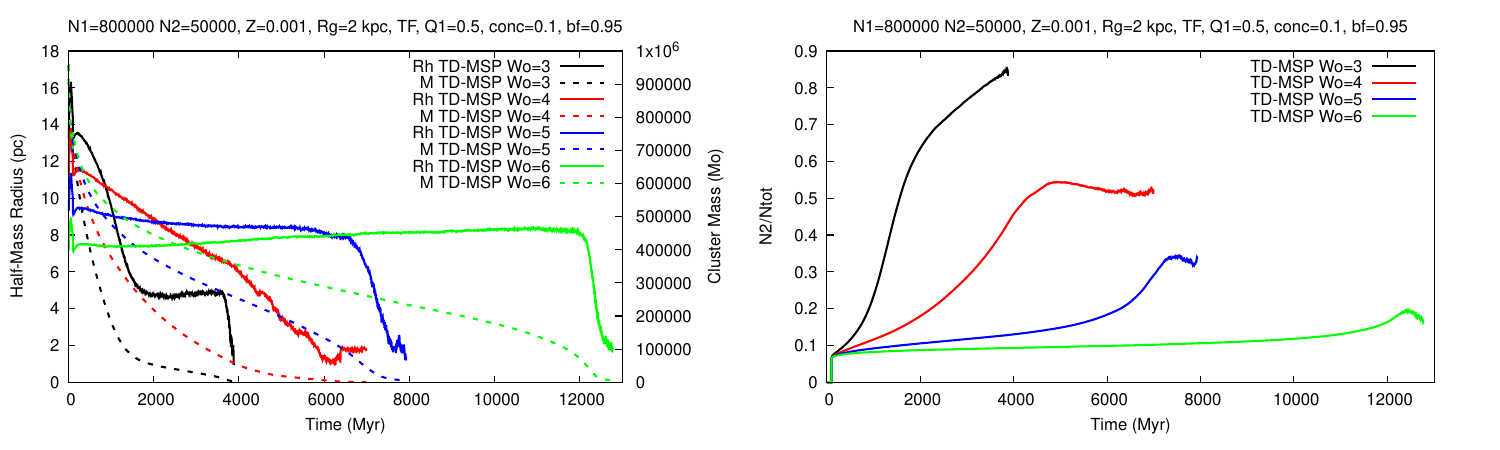}} 
    \caption{Left panel: Evolution of $\mathrm{R_h}$ (solid lines) and the cluster total mass (dashed lines) for TD-MSP models with a different King parameter, $\mathrm{W_{o1}}$. Right panel: Evolution of the ratio $\mathrm{N_2/N_{tot}}$ for TD-MSP models with a different King parameter, $\mathrm{W_{o1}}$.  The global cluster parameters are listed at the top of each panel and are the same as for Figure~\ref{f:Fig1}.}
    \label{f:Fig4} 
\end{figure*}

The works of \citet{Vesperinietal2021}, \citet{Hypkietal2022}, \citet{Hypkietal2025}, and \citet{Lacchinetal2024}\footnote{\citet{Lacchinetal2024} used a more sophisticated Galactic potential treatment and found strong mass loss even at $R_g$ = \rm{4} kpc, but for extremely filling 1P with $R_h$ = \rm{60} pc; and $R_t$ = \rm{200} pc, which confirms findings reported in the previous \mocca papers.} have clearly shown that for the AGB scenario of MSP formation, associated with gas re-accretion and ejected AGB envelopes, \fg must initially be TF and have a small concentration toward the center for the $\mathrm{N_2/N_{tot}}$ ratio to be comparable to the values observed in MWGCs. This condition for initial models described by the King parameter means that $\mathrm{W_{o1}}$ should be less than 6 or 5. Generally speaking, we expect that the smaller the $\mathrm{W_{o1}}$, the smaller the cluster mass, the larger the $\mathrm{R_h}$ after the gas re-accretion, and the larger the $\mathrm{N_2/N_{tot}}$ ratio. The further evolution is controlled by the escape speed. The larger the $\mathrm{W_{o1}}$, the higher the escape velocity and the smaller the stream of escaping objects from the cluster. Accordingly, the rapid loss of cluster mass and the associated decline in $\mathrm{R_t}$ should lead to a rapid decline in $\mathrm{R_h}$ and its relative stabilization when the BHS is responsible for the evolution of the cluster. 
Indeed, this type of behavior of global cluster parameters with increasing $\mathrm{W_{o1}}$ can be observed in Figure~\ref{f:Fig4}. There seems to be no monotonic dependence on $\mathrm{W_{o1}}$ for $\mathrm{R_h}$ evolution. It seems that for $\mathrm{W_{o1}}=5$ there is a trend reversal, for larger $\mathrm{W_{o1}}$, the level of almost constant $\mathrm{R_h}$ (associated with balanced BHS evolution) decreases instead of increasing. This is connected with a larger escape speed for larger $W_{o1}$ models. In addition, it seems that for $\mathrm{W_{o1}}=4$ the system does not show the characteristics of balanced evolution - $\mathrm{R_h}$ is continuously decreasing. This is an unexpected effect that requires further investigation. It seems that despite the presence of the BHS, the period of evolution in which cluster expansion is associated with the energy generated by binary systems, balances tidal stripping may not always be clearly marked.

\subsection{\sg concentration parameter}
\label{s:conc}
\begin{figure*}
    \centering
    \subfigure[]{\includegraphics[width=1.0\textwidth]{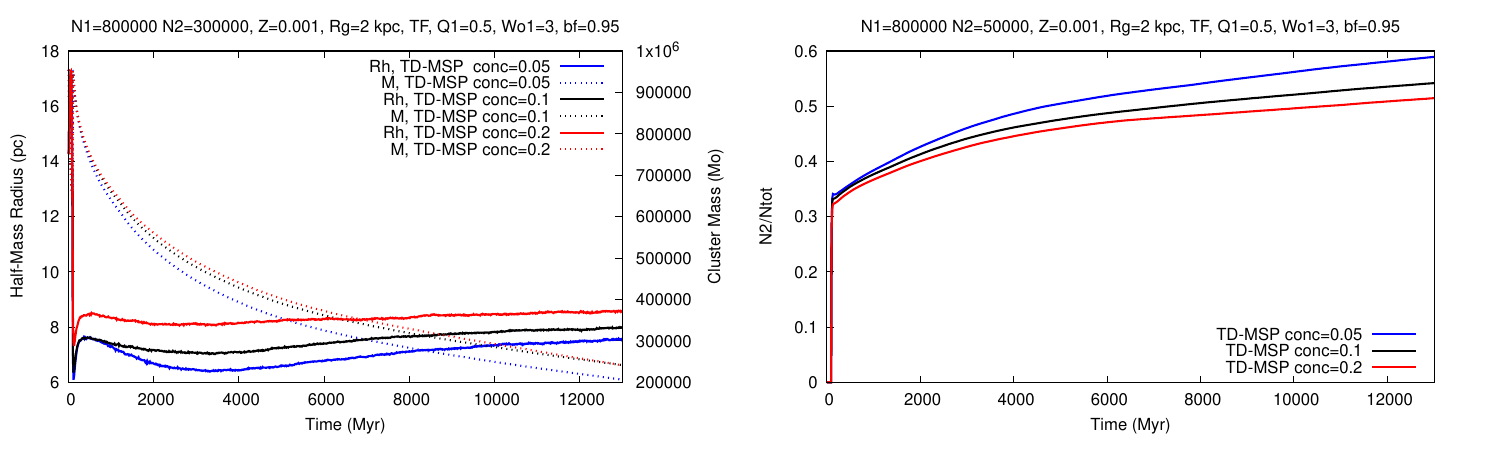}} 
    \caption{Left panel: Evolution of $\mathrm{R_h}$ (solid lines) and the cluster total mass (dashed lines) for TD-MSP models with different \sg concentration parameters ($\mathrm{conc=conc_{pop}}$). Right panel: Evolution of the ratio $\mathrm{N_2/N_{tot}}$ for TD-MSP models different $\mathrm{conc=conc_{pop}}$. The global cluster parameters are listed at the top of each panel and are the same as for Figure~\ref{f:Fig1}.}
    \label{f:Fig5} 
\end{figure*}

The concentration parameter ($\mathrm{conc_{pop}}$) is defined as the ratio of the half-mass radius of \sg to the half-mass radius of \fg. Thus, a cluster with a smaller $\mathrm{conc_{pop}}$ should be more concentrated, have a higher escape velocity, and become more nTF after \sg formation. This means that the larger the $\mathrm{conc_{pop}}$, the larger the mass of the cluster, the larger the $\mathrm{R_h}$, and the smaller the $\mathrm{N_2/N_{tot}}$ ratio. Indeed, in Figure~\ref{f:Fig5} we observe exactly such an evolution of GC global parameters depending on $\mathrm{conc_{pop}}$. It is noteworthy that the differences between the evolution of the cluster with different $\mathrm{conc_{pop}}$ do not result in a large change in the cluster's global parameters.

\subsection{Galactocentric distance}
\label{s:distance}
\begin{figure*}
    \centering
    \subfigure[]{\includegraphics[width=1.0\textwidth]{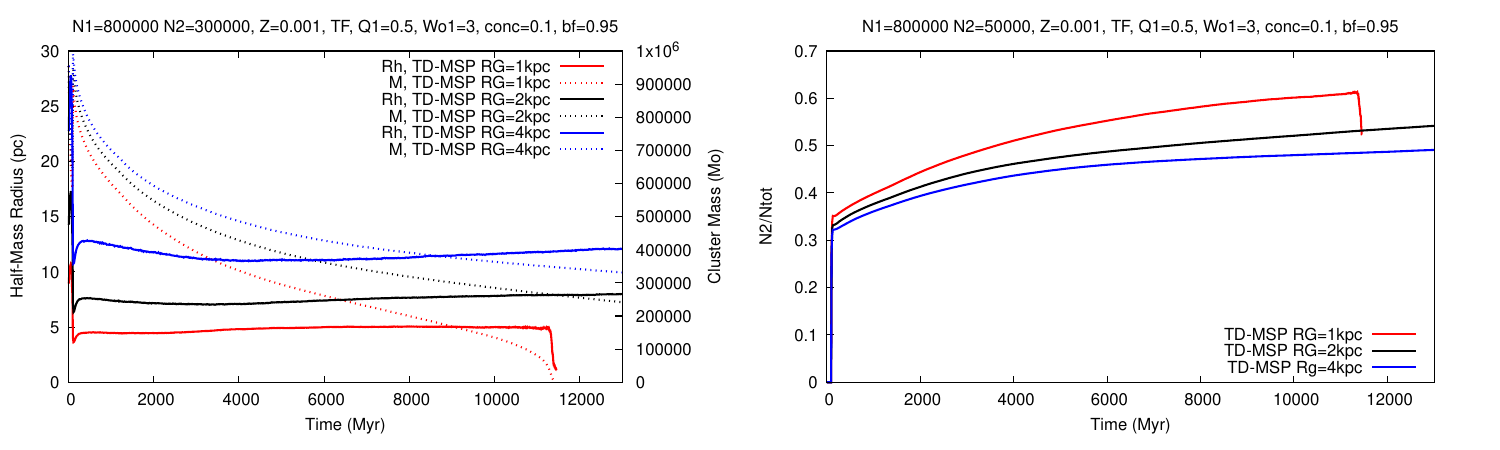}} 
    \caption{Left panel: Evolution of $\mathrm{R_h}$ (solid lines) and the cluster total mass (dashed lines) for TD-MSP models with different galactocentric distance ($\mathrm{R_g}$). Right panel: Evolution of the ratio $\mathrm{N_2/N_{tot}}$ for TD-MSP models different ${R_g}$. The global cluster parameters are listed at the top of each panel and are the same as for Figure~\ref{f:Fig1}.}
    \label{f:Fig6} 
\end{figure*}

The $\mathrm{R_t}$, assuming a constant rotational velocity of the galaxy and a circular orbit of the cluster, is just a function of the mass of the cluster and the $\mathrm{R_g}$. This means that, for the same cluster mass, the greater the $\mathrm{R_g}$, the greater the $\mathrm{R_t}$. Since the cluster is TF, this means that, for the same cluster spatial structure, the larger the $\mathrm{R_t}$, the larger the $\mathrm{R_h}$ and the half-mass relaxation time, and thus the slower is the dynamical evolution of the cluster. In summary, the larger the $\mathrm{R_g}$ the larger the mass of the cluster, the larger $\mathrm{R_h}$, and the smaller the $\mathrm{N_2/N_{tot}}$ ratio.
Indeed, we can observe this type of evolution of global cluster parameters in Figure~\ref{f:Fig6}. The cluster with the smallest $\mathrm{R_g}$ is characterized by the smallest $\mathrm{R_h}$, mass and evolution time, and the largest $\mathrm{N_2/N_{tot}}$ ratio. It is important to draw attention to the fact that dependence on $\mathrm{R_g}$ is important from the point of view of MWGC observational parameters. In the TD-MSP model, it is very difficult to obtain a range of MWGCs observational parameters, such as a relatively small $R_h$ or very large cluster masses. The solution, at least partially to this problem, is the formation of GCs very close to the center of the galaxy. This problem will be discussed in depth in Section~\ref{s:Discussion}.
As is seen in Fig.~\ref{f:Fig6}, the dependence of the $\mathrm{N_2/N_{tot}}$ ratio on the initial distance from the galactic center arises from the specific initial conditions adopted in our models. This trend is expected to be strongly blurred -- or even erased -- during the subsequent evolution of clusters due to environmental effects associated with galaxy assembly, GC orbital migration, and significant orbital eccentricities (see Section~\ref{s:Specul} for further discussion). In observed MWGCs, such a correlation is not found.
   
\subsection{\fg virial ratio}
\label{s:Q1}
\begin{figure*}
    \centering
    \subfigure[]{\includegraphics[width=1.0\textwidth]{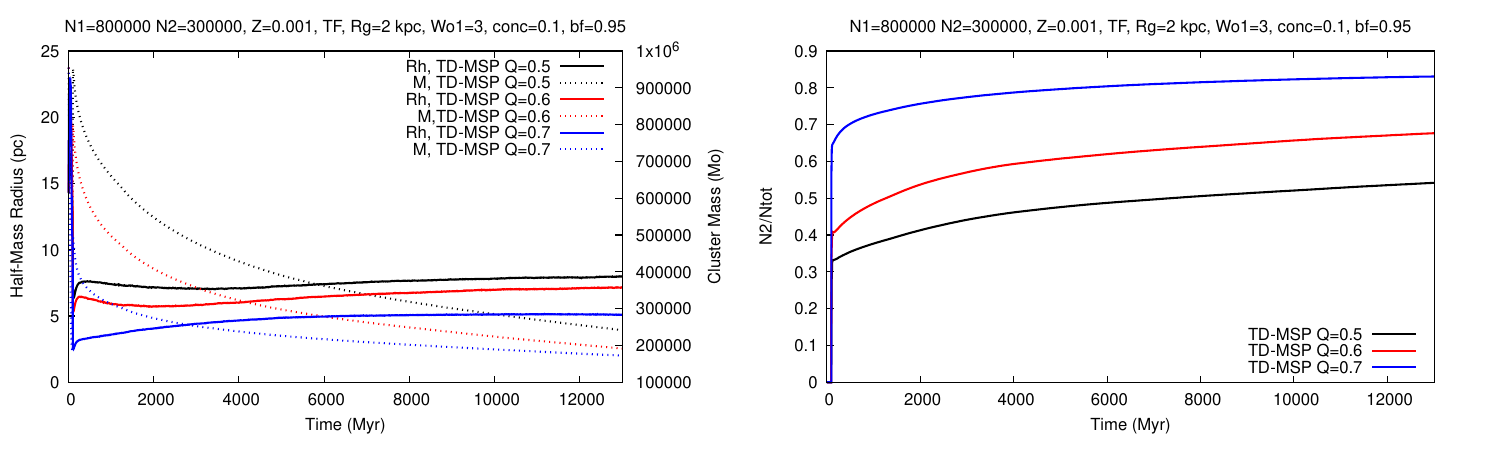}} 
    \caption{Left panel: Evolution of $\mathrm{R_h}$ (solid lines) and the cluster total mass (dashed lines) for TD-MSP models with different $\mathrm{Q_1}$. Right panel: Evolution of the ratio $\mathrm{N_2/N_{tot}}$ for TD-MSP models with different $\mathrm{Q_1}$. The global cluster parameters are listed at the top of each panel and are the same as for Figure~\ref{f:Fig1}.}
    \label{f:Fig7} 
\end{figure*}

The important parameter describing MSPs in the TD-MSP model is the virial ratio, $\mathrm{Q_1}$, for \fg. The process of star cluster formation is still a hot topic, not yet fully solved. The expansion of the cluster associated with the rejection of gas remaining after the formation of \fg was well established in theory and numerical simulations \citep[e.g.,][]{BanerjeeKroupa2013, BanerjeeKroupa2018, GoodwinBastian2006, Krauseetal2016}. However, according to recent works, the amount of cluster expansion due to the residual gas removal is very uncertain and depends on many factors connected with the star formation efficiency (SFE), cluster structure, environment, and efficiency of the feedback processes  \citep[e.g.,][]{Farias2018, Geenetal2018, Zamora-Avilesetal2019, Krauseetal2020, Polaketal2024}.
Therefore, the assumption that initially \fg is outside virial equilibrium and has $\mathrm{Q_1} > 0.5$ is fully natural. Overall, for systems with $\mathrm{Q_1} > 0.5$, we should expect stronger mass loss and tidal stripping, which will lead to lower cluster mass, lower $\mathrm{R_h}$, and a higher $\mathrm{N_2/N_{tot}}$ ratio. A model with a larger $\mathrm{Q_1}$ is characterized by an excess of kinetic energy relative to potential energy. Therefore, the spatial distribution of stars is less concentrated, and stars can escape from the system more easily. This leads to a sharp loss of \fg stars and a sharp decrease in $\mathrm{R_h}$ and a strong increase in the $\mathrm{N_2/N_{tot}}$ ratio. Indeed, this type of evolution is observed in Figure~\ref{f:Fig7}. The dependence of the evolution of the cluster's global parameters on $\mathrm{Q_1}$ seems crucial to reconstruct the range of global parameters observed in MWGCs. By taking into account the dependence on $\mathrm{Q_1}$, we can obtain $\mathrm{R_h}$ of several parsecs, and $\mathrm{N_2/N_{tot}}$ ratio greater than 0.8, but we cannot reproduce the observational relationship between cluster mass and $\mathrm{N_2/N_{tot}}$ - the greater the mass of the cluster, the greater the $\mathrm{N_2/N_{tot}}$ ratio. This problem and its possible solution are discussed in Section~\ref{s:Discussion}.

\subsection{Maximum IMF mass for \sg}
\label{s:IMF}
\begin{figure*}
    \centering
    \subfigure[]{\includegraphics[width=1.0\textwidth]{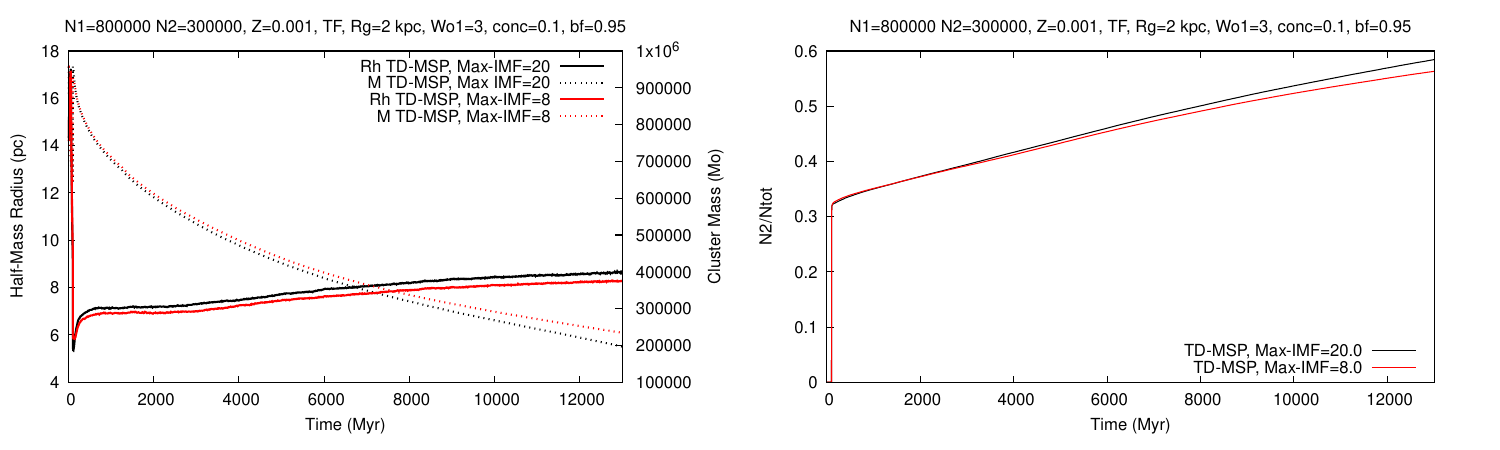}} 
\caption{Left panel: Evolution of $\mathrm{R_h}$ (solid lines) and the cluster total mass (dashed lines) for TD-MSP models with different maximum IMF mass for \sg objects. Right panel: Evolution of the ratio $\mathrm{N_2/N_{tot}}$ for TD-MSP models different maximum IMF mass for \sg objects.  The global cluster parameters are listed at the top of each panel and are the same as for Figure~\ref{f:Fig1}.}
    \label{f:Fig8} 
\end{figure*}

The maximum mass of stars formed in \sg is a quantity that is very difficult to determine within the framework of theory as well as observations. Therefore, it is usually assumed, for simplicity, that \sg is formed with the same IMF as \fg, only the maximum mass of main sequence stars is much smaller for \sg than for \fg. This limitation is related to the fact that for \sg, no significant differences in metal content are observed between \fg and \sg objects. This means that in \sg supernovae (SNe) outbursts should be kept to a minimum in order not to contaminate with metals the gas from which the less massive \sg stars are formed. By reducing the maximum mass for IMF from 20 $\mathrm{M_{\odot}}$ to 8 $\mathrm{M_{\odot}}$ \citep[e.g.,][]{D'Ercoleetal2008, Bekki2019, Sollimaetal2022}, we should expect a smaller reduction in $\mathrm{R_h}$ as a result of gas re-accretion and a smaller cluster mass and a lower $\mathrm{N_2/N_{tot}}$ ratio. This is because a cluster with a smaller IMF maximum mass is less massive than a cluster with a larger IMF maximum mass (with the same IMF). Thus, a cluster with a larger maximum IMF mass has a larger $\mathrm{R_t}$ and thus $\mathrm{R_h}$. Larger $\mathrm{R_h}$ also means a little more mass loss by the cluster and consequently larger $\mathrm{N_2/N_{tot}}$. Indeed, we can observe such behavior in Figure~\ref{f:Fig8}, which shows the evolution of the cluster mass, $\mathrm{R_h}$, and the $\mathrm{N_2/N_{tot}}$ ratio as a function of the maximum IMF mass. Changing the maximum IMF mass has only a small effect on the evolution of these parameters. Of course, the number of BHs or NSs will change and will be slightly smaller (about 5\%) for a maximum IMF mass of 8 $\mathrm{M_{\odot}}$. Such small changes do not have a strong influence on the GC evolution. The above results suggest that the maximum IMF mass does not have a key effect on the global observational parameters of GCs, provided that it is not too large and affects the metallicity of \sg, or too small to allow the survival of GCs that are TF and have a non-concentrated \fg with small $W_{o1}$.

\subsection{Time of gas re-accretion}
\label{s:Time}
\begin{figure*}
    \centering
    \subfigure[]{\includegraphics[width=1.0\textwidth]{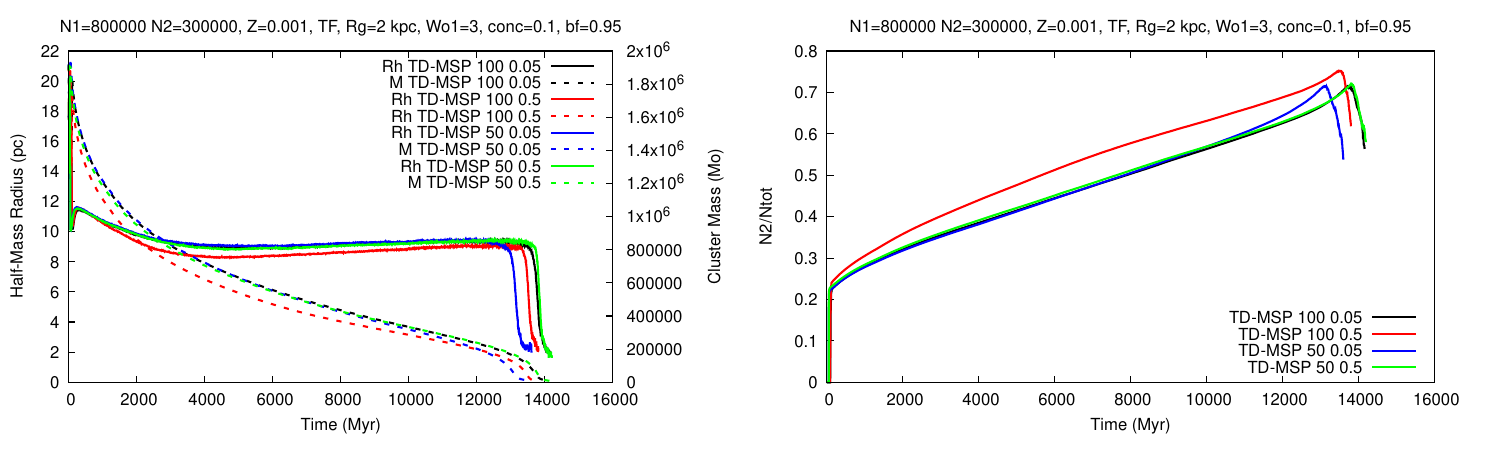}} 
\caption{Left panel: Evolution of $\mathrm{R_h}$ (solid lines) and the cluster total mass (dashed lines) for TD-MSP models with different time of gas re-accretion (combination of $\mathrm{t_{delay} = t_{end}}$ time (100 or 50 Myr) and the fraction of $\mathrm{t_{delay}}$ time at which the gas re-accretion starts ($\mathrm{t_{start}}$: 0,05 or 0.5))  Right panel: Evolution of the ratio $\mathrm{N_2/N_{tot}}$ for TD-MSP models different time of gas re-accretion. The global cluster parameters are listed at the top of each panel and are the same as for Figure~\ref{f:Fig1}.}
    \label{f:Fig9} 
\end{figure*}

The implementation of the AGB scenario for MSP formation in the \mocca code is based on several technical parameters, in particular, those describing the time of the start of gas re-accretion ($\mathrm{t_{start}}$) and the time of \sg formation ($\mathrm{t_{delay}}$ - see Table~\ref{t:Tab1}). These parameters affect the scale of cluster expansion associated with mass loss due to stellar evolution and the scale of $\mathrm{R_h}$ decrease. The earlier the start of gas re-accretion, the smaller the initial $\mathrm{R_h}$ expansion and the smaller the mass loss of the cluster. Similarly, the later the time of \sg formation, the greater the expansion of $\mathrm{R_h}$ and the greater the cluster's mass loss. Since the mass of re-accreted gas is independent of the technical parameters, we can expect that the magnitude of the reduction in $\mathrm{R_h}$ and the increase in cluster mass after \sg formation will be similar to the values obtained in the reference model. We can only expect minor differences related to an earlier or later start of the evolution of \sg objects, and a slightly reduced rate of mass loss by \fg. Thus, the observational parameters of GCs discussed in the paper should not show a significant dependence on the technical parameters describing gas re-accretion. Indeed, these predictions are confirmed by the simulations shown in Figure~\ref{f:Fig9}. The long-scale evolution of GCs should not significantly depend on the time of the start of gas re-accretion and the time of \sg formation. Other model parameters related to the structure of \fg and \sg and the environment in which the cluster moves are of key importance and provide an opportunity to estimate these parameters by comparing models with observations.

\subsection{Brief summary}
\label{s:Brief}
In two previous papers \citep[][]{Hypkietal2022, Hypkietal2025}, it was shown that under the nTD-MSP scenario, the evolution of GCs with MSPs can successfully reproduce the observational parameter ranges of MWGCs. However, a very serious flaw in this model was the assumption that both populations arise simultaneously and both are in virial equilibrium. To break free from this assumption, the TD-MSP model was introduced, in which the formation of \sg arises at a certain time after \fg. In addition, GCs can migrate to greater galactocentric distances after a certain evolutionary time. The new models are characterized by many parameters described in Table~\ref{t:Tab1}. Unfortunately, the introduction of the TD-MSP model has resulted in global model parameters that less closely reproduce the observational parameters of MWGCs; in particular, the $\mathrm{N_2/N_{tot}}$ ratio is clearly smaller, $\mathrm{R_h}$ rises to much higher values, and cluster mass is smaller than that of nTD-MSPs. This is very evident for models with migration. 

Of course, by manipulating the global and technical parameters describing MSPs, one can try to get model parameters more in line with MWGCs, i.e., $\mathrm{R_h}$ within a few parsecs, $\mathrm{N_2/N_{tot}}$ within 0.3 to 0.9, and masses up to several hundred thousand $\mathrm{M_{\odot}}$. The most promising factors are:

\begin{enumerate}
\item Number of \sg objects ($\mathrm{N_2}$): A smaller $\mathrm{N_2}$ leads to a higher $\mathrm{N_2/N_{tot}}$ ratio and a smaller half-mass radius ($\mathrm{R_h}$).  Unfortunately, the final mass of the GC is too small.
\item Number of \fg objects: A larger number of \fg objects slows cluster evolution. When combined with a small number of \sg objects, this can support cluster survival until Hubble time.
\item Central concentration of \fg ($\mathrm{W_{o1}}$): A smaller $\mathrm{W_{o1}}$ results in a smaller $\mathrm{R_h}$ and cluster mass, a higher $\mathrm{N_2/N_{tot}}$ ratio, but also leads to significantly faster cluster evolution.
\item Galactocentric distance ($\mathrm{R_g}$): A smaller $\mathrm{R_g}$ leads to a smaller $\mathrm{R_h}$ and cluster mass, a higher $\mathrm{N_2/N_{tot}}$ ratio, but also results in faster cluster evolution.
\item Virial parameter ($\mathrm{Q_1}$): A larger $\mathrm{Q_1}$ results in a smaller $\mathrm{R_h}$ and cluster mass, a higher $\mathrm{N_2/N_{tot}}$ ratio, but also leads to significantly faster cluster evolution.
\end{enumerate}

It seems that $\mathrm{Q_1}$ is the most promising parameter to obtain the observational values of $\mathrm{N_2/N_{tot}}$ and $\mathrm{R_h}$ in the TD-MSP models. Unfortunately, the combination of this parameter with other parameters will not allow for cluster masses to be obtained at their upper observable values. The solution may be another mechanism related to the environment in which GCs live. In the next section, we discuss possible solutions to this problem.

\section{Discussions}
\label{s:Discussion}
This section focuses on briefly summarizing the observational material collected mainly photometrically (\mocca simulations only with this type of observation can be compared) for GCs with MSPs. We then provide a brief description of the scenarios proposed to explain MSPs, discuss the weaknesses, and identify the question marks associated with these scenarios. We then present a speculative refinement of the AGB scenario for the emergence and evolution of MSPs. 

\subsection{Observations}
\label{s:Obs}
The observational characteristics of GCs with MSPs are very extensively and critically discussed in the review paper by \citet{Bastian2018}. Briefly, the following facts about MSPs in GCs have solid observational confirmation.
\begin{enumerate}
\item Massive GCs are composed of stars exhibiting differences in light element content. Correlations and anticorrelations between the contents of these elements indicate that they must have been formed by thermonuclear reactions of hydrogen burning at very high temperatures, which can happen only in the case of massive stars \citep[e.g.,][]{Denisenkov1989, Langeretal1993, Prantzosetal2007};
\item MSPs are characterized by a very small range of Fe abundances (only about 0.1 dex);
\item Differences in the ages of different populations are very small, only on the order of a few tens of millions of years;
\item \sg is centrally concentrated relative to \fg \citep[e.g.,][]{Bastian2018, MastrobuonoBattisti2021}, however, see the recent work of \citet{Leitingeretal2023};
\item MSPs are observed in different galactic environments. Clusters with too low masses or too young do not seem to show MSPs. The critical mass seems to be around $10^5$ $\mathrm{M_{\odot}}$ and age smaller than 1 – 2 Gyr \citep[][]{Miloneetal2020, Leitingeretal2023};
\item There is a relatively strong correlation between GC mass and $\mathrm{N_2/N_{tot}}$ ratio \citep[e.g.,][]{Bastian2018};
\item $\mathrm{N_2/N_{tot}}$ ratio does not seem to correlate clearly with other global parameters of GCs.
\end{enumerate}
Any scenario attempting to explain the formation of MSPs in GCs must take into account the above observational facts. An additional complication is the fact of the violent process of MW formation associated with multiple close interactions and mergers with nearby dwarf galaxies. In the course of this, many GCs associated with dwarf galaxies have been intercepted and accreted by the MW \citep[e.g.,][]{Chen2024}. The population of MWGCs is a collection of GCs formed in different environments, and this fact must be taken into account when we try to understand the observational facts of MSPs and compare them with simulations of the evolution of GCs. 

As was mentioned earlier, the MOCCA code can only be used to study the evolution of clusters in circular orbits around a point mass, with a mass equal to that of the galaxy contained within the orbit. This assumption is at odds with the orbits of MWGCs \citep[e.g.,][]{Bajkova2021}. In order to be able to compare the results of simulations with observations, we used the work of \citet{Caietal2016} in which the authors define a circular orbit, which is characterized by a similar cluster mass loss as the cluster being in an eccentric orbit.  The relationship between circular orbit and eccentric orbit parameters is as follows, $\mathrm{R_c} = \mathrm{a}(1+\mathrm{e})(1-0.71\mathrm{e})^{(5/3)}$, where $\mathrm{R_c}$ is the radius of the circular orbit, and a and e are the semimajor axis and eccentricity, respectively. Of course, the formula based on Table 1 in \citet{Caietal2016} is very approximate, but nevertheless, it gives some insight into the rate of evolution of the MWGC, which can be compared with that obtained in \mocca simulations. Note that mass loss of GCs is the main factor responsible for the evolution of the $\mathrm{N_2/N_{tot}}$ ratio. Thus, a reasonably accurate way to establish the relationship between the actual and simulated evolution of GCs in numerical models is crucial to better understand the observations of MWGCs. Recently, \citet{Chen2024} analyzed 10 structural and kinematic observational parameters of MWGCs to determine which GCs formed in situ and which ex situ. They estimated that about 60\% of the MWGCs were formed in situ. Using their results, we shall present figures defining the relationship of the observational parameters with the type of their formation.

\begin{figure*}
    \centering
    \subfigure[]{\includegraphics[width=1.0\textwidth]{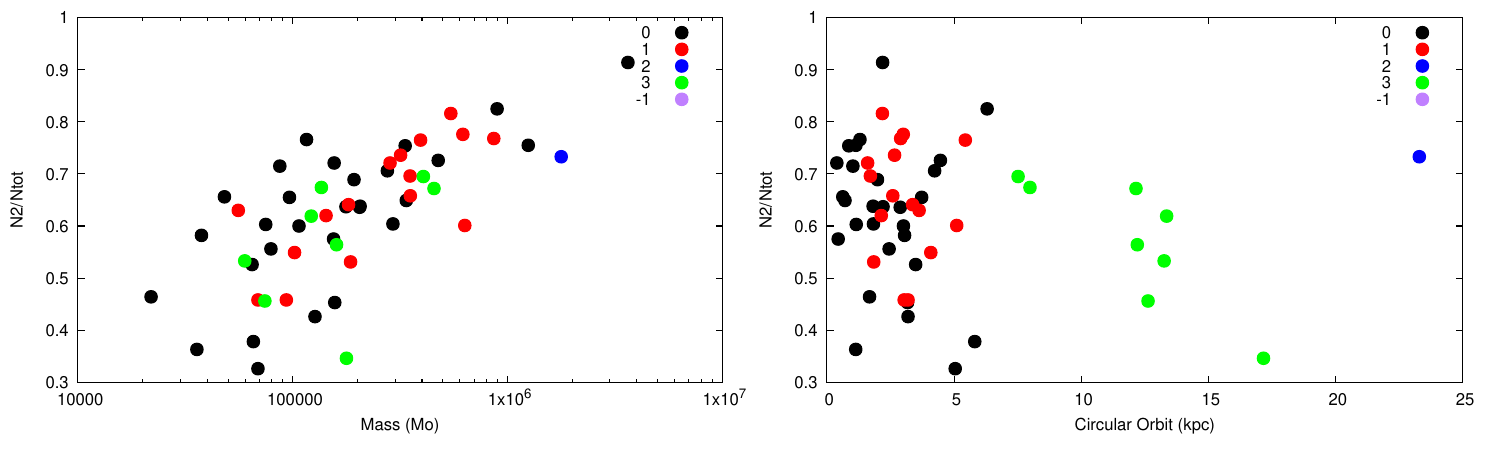}} 
    \caption{Left panel: Dependence of the $\mathrm{N_2/N_{tot}}$ ratio on the present-day mass of MWGCs. Right panel: $\mathrm{N_2/N_{tot}}$ ratio vs. the radius of the circular orbit. Colored dots indicate GC type, whether it formed in the MW or arrived with a dwarf galaxy absorbed by the MW: 0, black - in situ formation, 1, red - ex situ Gaia-Sausage/Enceladus formation, 2, blue - ex situ Sagittarius dwarf, 3, green - ex situ other mergers with dwarf galaxies, -1, purple - type not specified. Data are from catalogs: \citet{Miloneetal2017, Bajkova2021, Chen2024, Baumgardtetal2019}.}
    \label{f:Fig10} 
\end{figure*}

\begin{figure*}
    \centering
    \subfigure[]{\includegraphics[width=1.0\textwidth]{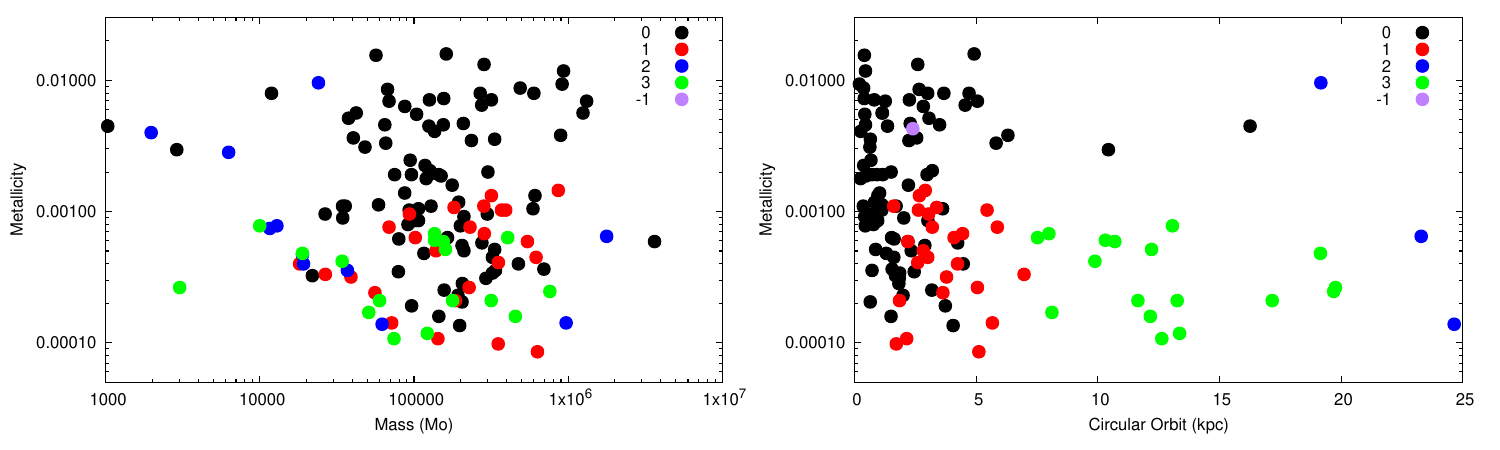}} 
    \caption{Left panel: Cluster metallicity ($\mathrm{Z}$) vs. present-day MWGC mass. Right panel: Dependence of the cluster metallicity on the size of the circular orbit. A description of the color bar is given in the caption of Figure~\ref{f:Fig10}.}
    \label{f:Fig11} 
\end{figure*}

\begin{figure*}
    \centering
    \subfigure[]{\includegraphics[width=1.0\textwidth]{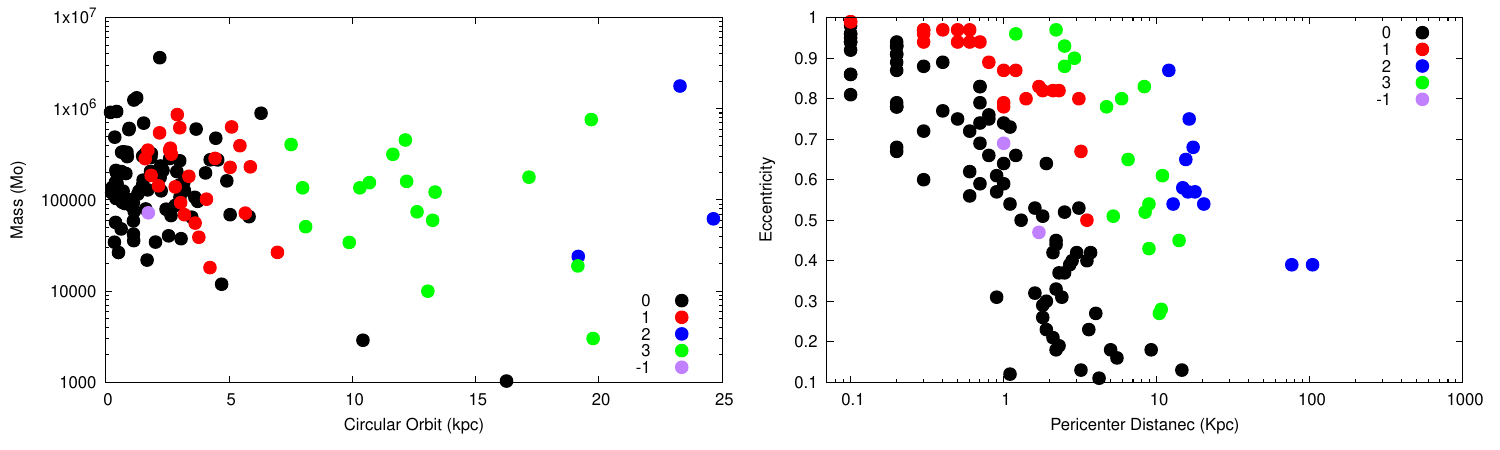}} 
    \caption{Left panel: Dependence of the cluster mass on the radius of the circular orbit. Right panel: Dependence of the cluster orbital eccentricity on the orbit pericenter distance. A description of the color bar is given in the caption of Figure~\ref{f:Fig10}.}
    \label{f:Fig12} 
\end{figure*}

The observational data shown in Figures~\ref{f:Fig10}--\ref{f:Fig12} are from the following MWGC catalogs: \citet{Miloneetal2017} for $\mathrm{N_2/N_{tot}}$ ratios, \citet{Bajkova2021} for kinematic parameters, \citet{Chen2024} for origin types, and \citet{Baumgardtetal2019} for global parameters. Figure~\ref{f:Fig10} shows $\mathrm{N_2/N_{tot}}$ ratios as a function of cluster mass (left panel) and radius of the circular orbit (right panel). In the left panel, we can clearly see, widely discussed in the literature, the correlation between $\mathrm{N_2/N_{tot}}$ and cluster mass \citep[e.g.,][]{Carrettaetal2010, Milone2015, Miloneetal2017}. The correlation is relatively strong and, interestingly, shows no significant differences between clusters formed in situ and ex situ.
\citet{Yaghoobietal2022},  using hydrodynamic simulations of 2P formation in young massive clusters, found that dense clusters can exhibit a positive correlation between the 2P fraction and cluster mass immediately after 2P formation. In their models, the 2P fraction ranges from approximately 2\% to 30\%, values comparable to the initial values assumed in our simulations. However, present-day GCs typically show higher enriched fractions, indicating that further dynamical evolution must play a significant role. In our models, this evolution is shaped by environmental effects such as the cluster’s migration through the galaxy and the strength of the tidal field. As discussed in Section \ref{s:Specul}, we propose a scenario in which more massive GCs, forming closer to the galactic center, re-accrete more gas and experience stronger tidal fields, leading to an increase in the 2P fraction over time. This scenario complements the initial trends found by \citet{Yaghoobietal2022} and provides a pathway for connecting early cluster conditions to present-day observations.
It seems that this type of correlation could indicate the effect of the tidal field of the parent galaxy. Tidal field forces are similar, which means that GCs in dwarf galaxies should form closer to the center than in the MW. The right panel reveals an interesting (maybe expected) dependence on the radius of the circular orbit. Ex situ GCs have average orbit radii significantly larger than in situ clusters. In general, most in situ clusters have radii of circular orbits contained within 7 kpc (see also Figures~\ref{f:Fig11} and \ref{f:Fig12}). 

Figure~\ref{f:Fig11} shows cluster metallicity as a function of cluster mass (left panel) and the radius of the circular orbit (right panel). In the right panel, we can see that in general there is no obvious relationship between cluster mass and metallicity, except for the fact that ex situ clusters show a smaller spread of metallicity, and a large fraction of in situ clusters have higher metallicity than ex situ clusters. In the right panel, we see a very clear dependence on the radius of the circular orbit. Ex situ clusters have, on average, a lower metallicity than in situ clusters and are distributed over much wider orbits than in situ clusters. This is even more pronounced than in Figure~\ref{f:Fig10}.

Figure~\ref{f:Fig12} shows the mass of the cluster as a function of the radius of the circular orbit (left panel) and the eccentricity of the cluster's orbit as a function of pericenter distance (right panel). As we expected after analyzing the previous figures, there is no dependence of the cluster mass on the radius of the circular orbit. Ex situ GCs have a similar mass distribution to in situ clusters.  In contrast, in the right panel, we see an interesting correlation between eccentricity, distance at the pericenter, and GC type. Clusters with orbits of lower eccentricity have significantly larger distances at the pericenter than clusters with higher eccentricities. Their orbits are more circular. It is also interesting that ex situ clusters show a lack of close circular orbits and have distances at the pericenter shifted to greater distances than in situ clusters. This is an indication that they arrived in the MW along with dwarf galaxies.

In conclusion, there seems to be no strong observational data indicating the dependence of the $\mathrm{N_2/N_{tot}}$ - cluster mass correlation on the type of environment where GCs formed. This tells us, in our opinion, that the strength of tidal interactions of the just-forming galaxy is crucial in building the above correlation. As Figure~\ref{f:Fig10} shows, the maximum mass of GCs does not seem to depend on whether they formed in situ or ex situ. This may suggest that the formation process of GCs is universal and depends on the strength of the tidal field and on the availability of gas. Ex situ GCs have lower metallicity and are distributed in wider orbits with greater distance at the pericenter than in situ GCs.  Also, the ex situ population lacks orbits with low eccentricity, which may be related to events of close MW interactions with dwarf galaxies. 

\begin{figure*}
    \centering
    \subfigure[]{\includegraphics[width=1.0\textwidth]{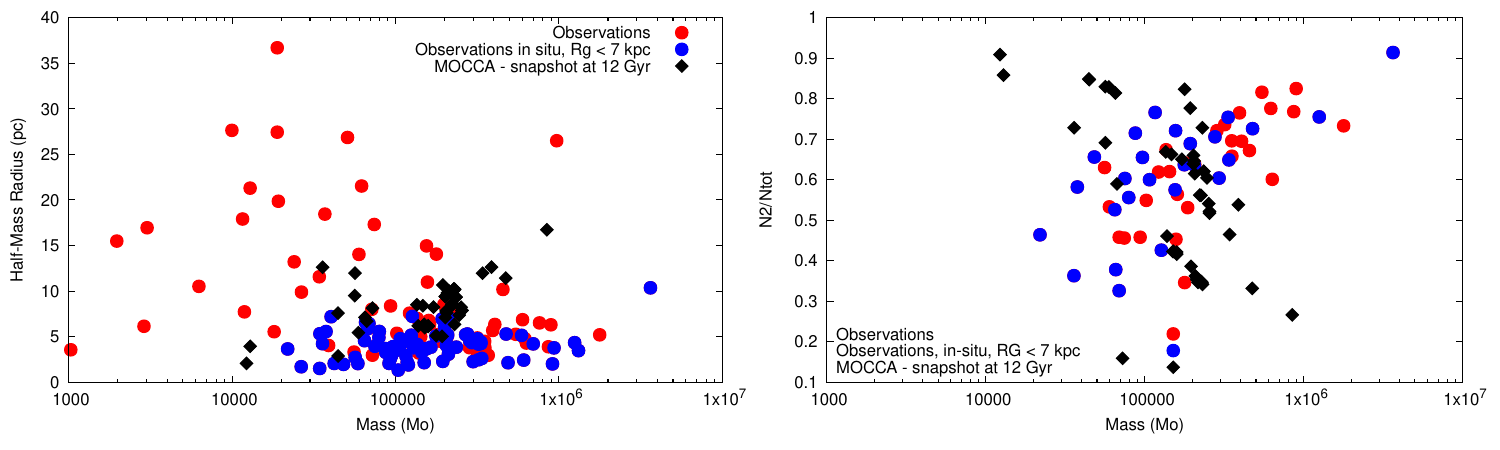}} 
    \caption{Comparison between observational parameters of MWGCs and simulation results. Red dots represent the observations, blue dots the observations limited to in situ MWGCs and $\mathrm{R_g}$ smaller than 7 kpc, and black diamonds the simulations. Left panel – relation between cluster mass and $\mathrm{R_h}$. Right panel: Relation between cluster mass and $\mathrm{N_2/N_{tot}}$ ratio. Please note that the anticorrelation relationship between cluster mass and $\mathrm{N_2/N_{tot}}$ ratio observed in \mocca models is purely artificial and is related to the choice of initial models. The nTF models and those initially located at large galactocentric distances cannot reproduce the observed parameters of MWGCs. }
    \label{f:Fig13} 
\end{figure*}

Figure~\ref{f:Fig13} shows the comparison between observational data for MWGCs for all cluster populations and part of the population restricted to the in situ GCs and with galactocentric distances smaller than 7 kpc, and simulation results presented in this paper.
While our initial model parameters were chosen to explore the impact of specific assumptions related to MSP formation and evolution -- rather than to reproduce the full diversity of the MWGC population -- the agreement between the simulation results and observations is generally satisfactory, particularly in terms of the $N_2/N_{\mathrm{tot}}$ values.
However, we acknowledge that the simulated clusters exhibit an anticorrelation between $N_2/N_{tot}$ and present-day cluster mass, whereas observational data suggest a positive or flat correlation. This discrepancy will be further discussed in Section 4.3. One possible explanation for this difference is the assumption of a static galactic potential in our simulations, which may not fully capture the early evolution of clusters in a time-dependent tidal field. Accounting for variations in the galactic potential, as suggested by \citet{Meng2022}, could provide a more accurate representation of cluster evolution and improve agreement with observations. In particular, the parameters of MWGCs in situ and located relatively close to the MW center are reproduced satisfactorily in the simulations. Of course, as it was mentioned earlier $\mathrm{R_h}$ for models is a bit too large, and models with masses close to $10^6 \mathrm{M_{\odot}}$ are missing. Using the results presented in Section~\ref{s:Results}, it will be possible to select the initial parameters of the models to best match the observational parameters of the MWGCs.

In summary, to compare \mocca simulations with observational data, we should mainly focus on in situ GCs with the radii of circular orbits below about 7 - 10 kpc. As was already mentioned in Section~\ref{s:Method}, the \mocca code explicitly assumes a tidal field of the present MW. Taking into account the tidal fields of different galaxy types, particularly during the early stages of GC evolution, would require simplified models that incorporate key insights from cosmological simulations.

\subsection{MSP formation scenarios}
\label{s:Scenarios}
Several scenarios have been proposed to explain the formation of MSPs in GCs, but none fully account for all observed properties \citep[e.g.,][]{Bastian2018}. The most widely discussed mechanisms include:

\begin{enumerate}
\item The AGB scenario \citep[e.g.,][]{DErcoleetal2008, Bekki2017, Caluraetal2019}
\item Fast-Rotating Massive Stars and Interacting Binaries (FRMS) \citep[e.g.,][]{Decressinetal2007}
\item The Early Disk Accretion Scenario \citep[e.g.,][]{Bastianetal2013}
\item Very Massive Stars Due to Runaway Collisions (VMS) \citep[e.g.,][]{Gielesetal2018} 
\end{enumerate}

We briefly summarize these models and their challenges in explaining MSP formation, particularly regarding the required initial conditions, gas retention, and kinematic properties.

\subsubsection{AGB scenario}
\label{s:AGB}
The AGB scenario has already been briefly discussed in the previous sections. It has been shown that the proposed TD-MSP scenario can explain most of the observed parameters of MWGCs, except the correlation between $\mathrm{N_2/N_{tot}}$ ratio and the cluster present-day mass,  providing a more open view of the initial conditions and the environment in which GCs live. This is about the migration of GCs, and the initial model not being in virial equilibrium. Therefore, we shall now focus mainly on discussing the weaknesses of this scenario. It is important to note that there is no mass budget problem in this scenario. It is solved by assuming that \sg is more centrally concentrated and the cluster is TF or close to TF and will lose a significant fraction of \fg stars \citep[e.g.,][]{Vesperinietal2021, Hypkietal2022, Hypkietal2025}.

In our opinion, the biggest weakness is the \sg star formation process in a relatively dense gas and stellar environment. The question is what physical conditions must be met for \sg stars to form effectively in the dense environment of relatively massive and bright, mass-losing \fg stars, and what the efficiency is of this process. Some of these questions were addressed a relatively long time ago by \citet{Conroy2011}. They pointed out the importance of ram pressure and accretion from the ambient interstellar medium on the development of young GCs and type II Supernovae and prompt type Ia Supernovae. They also provided some constraints on the initial properties of GCs and the early galactic environment.
Unfortunately, their work was not followed by appropriate numerical simulations. The MOCCA models assume a \rm{100}\% SFE, which is supported by recent zoom-in cosmological simulations carried out, for example, by \citet{BrownGnedin2022, Caluraetal2024}, and simulations of isolated star-forming clouds by \citet{Polaketal2024}. The authors showed that SFE in dense and massive clusters can be as high as about \rm{85}\%. Such a high SFE for 2P would help avoid the mass budget problem for 2P, in which the mass of the re-accreted gas would be comparable to that of 1P. Such a large SFE for POP1 suggests a much-reduced cluster expansion after the residual gas has been ejected. 
Sustaining the large loss of 1P stars, in this case, will probably require that clusters form close to the galactic center and then migrate to larger distances from the center.
Interestingly, models that start with a lower $\mathrm{N_2/N_{tot}}$ ratio can evolve to have higher final fractions than models that were initially more enriched.
In these models, a slow increase in the $\mathrm{N_2/N_{tot}}$ ratio is observed first, which then accelerates significantly after 1-2 Gyr and brings the ratio to the observed values, which is consistent with the absence of \sg in clusters younger than 2 Gyr \citep[e.g.,][]{Mucciarellietal2007, Saracinoetal2020}.

\subsubsection{Alternative formation mechanisms for multiple stellar populations}\label{s4:alternatives}

Other proposed MSP formation mechanisms, including FRMS, early disk accretion, and VMS, face common challenges in explaining observations. In the FRMS scenario, rotational mixing in massive stars brings enriched material to the surface, where it is expelled through stellar winds and mixed with residual gas.
However, observations show that most young clusters expel gas within a few million years \citep[e.g.,][]{Bastianetal2014}, leaving little time for \sg\ star formation. As a result, the SFE challenges discussed in the AGB scenario also apply here. Simulations of single-population clusters indicate that reproducing the observed GC parameters requires initial conditions that strongly favor nTF clusters \citep{Askaretal2017}.
This implies that even after gas expulsion following \fg formation, the cluster remains nTF. In MSPs hosting nTF clusters, the initial $\mathrm{N_2/N_{tot}}$ ratio remains nearly constant throughout cluster evolution \citep{Hypkietal2022, Hypkietal2025}. Therefore, to account for the high fraction of \sg\ stars in present-day GCs, this scenario would require the total mass of \sg stars to exceed that of \fg stars, which is not plausible. Additionally, in dense nTF clusters, massive stars segregate to the center within a few million years \citep{ArcaSeddaetal2024}, causing residual gas to mix primarily in the core, leading to strongly centrally concentrated \sg\ formation. However, observations \citep{Bastian2018} show that the $\mathrm{N_2/N_{tot}}$ ratio remains nearly constant beyond $\mathrm{R_h}$, indicating that \sg\ stars are also present at large radii. 

The early disk accretion scenario suggests that low-mass stars retain protoplanetary discs for $\sim$10 Myr, sweeping up enriched material from massive star ejecta. However, cluster formation is rapid, and sub-clumps merging into a single cluster are unlikely to be significantly polluted before the final assembly phase. Additionally, the dense cluster environment raises doubts about whether disk accretion alone can generate the observed abundance variations.

The VMS scenario proposes that runaway stellar collisions in a dense cluster lead to the formation of a massive ($\gtrsim 1000 M_{\odot}$) star, which ejects H-burning material into the intracluster medium. However, VMS have short lifetimes ($\sim$3–4 Myr) and either collapse into an intermediate-mass BH or disrupt due to winds or instabilities \citep[e.g.,][]{PortegiesZwartetal2004, Gierszetal2015}. Runaway collisions require extreme densities, implying a strong nTF cluster and raising the same concerns as other formation scenarios. Moreover, intense VMS winds and radiation could expel processed material rather than allow it to mix with pristine gas. As in the FRMS scenario, VMSs are expected to segregate rapidly to the cluster center, causing enriched material to concentrate there, contradicting the observed widespread distribution of \sg\ stars \citep[e.g.,][]{Bastian2018, Hypkietal2025}. A hybrid approach incorporating disk accretion of processed material \citep[e.g.,][]{Bastianetal2013} might mitigate this issue, but it remains unclear whether enough material would be retained within the cluster.

Overall, these models struggle to explain MSP formation while remaining consistent with MWGC observations. In the next section, we explore a speculative AGB-based scenario informed by our results and observational data.

\subsection{Speculations}
\label{s:Specul}
Our speculative refinement of the AGB scenario on the formation and evolution of GCs with MSPs will be based on the results of simulations performed with the \mocca code and observational data, mainly kinematic and photometric. The speculations are rooted in the AGB model and will not attempt to explain the details of the observed correlations and anticorrelations between chemical elements. This will require detailed research beyond the scope of this paper. The focus is on trying to understand the global properties of MWGCs, namely the observed $\mathrm{R_h}$, $\mathrm{M}$, and $\mathrm{N_2/N_{tot}}$ ranges. 

The masses of MWGCs are contained in the range of roughly $10^4-10^6 \mathrm{M_{\odot}}$, $\mathrm{R_h}$ in the range of about 1 to 10 pc (larger radii have mostly ex situ GCs), and the $\mathrm{N_2/N_{tot}}$ ratio in the range of about 0.3 to 0.9. An exceptionally strong constraint on models of MSP formation in GCs is the strong correlation between the mass of the cluster and the $\mathrm{N_2/N_{tot}}$ ratio. Simulations provide us with the following suggestions (when we talk about observational parameters, they are always compared with simulations in which the age of the cluster is 12 Gyr):
\begin{enumerate}
\item GCs just after the expulsion of the gas left over from \fg formation should be TF or only slightly nTF. This means that they should form close to the galactic center. Too large galactocentric distances (keeping cluster mass constant) mean much larger $\mathrm{R_t}$. It is difficult to imagine that a cluster with $\mathrm{R_t}$ on the order of several hundred parsecs would form as a TF cluster;
\item Only models in which the GC is TF or only slightly TF allow us to achieve a $\mathrm{N_2/N_{tot}}$ ratio within the limits observed for MWGCs. The fact that the model is TF does not guarantee that the correlation between the $\mathrm{N_2/N_{tot}}$ ratio and the cluster mass will be reproduced;
\item According to recent zoom-in cosmological star cluster formation simulations, the SFE is very high for massive and dense GCs and can reach more than about \rm{85}\% \citep[][]{Caluraetal2024}. This means that the expansion of the cluster after the rejection of residual gas will be rather modest. Thus, for the $\mathrm{N_2/N_{tot}}$ ratio to be within the observed limits, according to the above points, the cluster has to form relatively close to the galactic center;
\item Increasing the galactocentric distance for newly formed GCs with the same masses increases their $\mathrm{R_h}$ and decreases the $\mathrm{N_2/N_{tot}}$ ratio. This follows from the assumption that clusters are TF;
\item Models with smaller $\mathrm{W_{o1}}$ guarantee larger values of the $\mathrm{N_2/N_{tot}}$ ratio. The larger the $\mathrm{W_{o1}}$ the greater the cluster mass, $\mathrm{R_h}$ and smaller the $\mathrm{N_2/N_{tot}}$ ratio;
\item The further a cluster is from virial equilibrium, the greater the $\mathrm{N_2/N_{tot}}$ ratio. The larger the $\mathrm{Q_1}$ the smaller the cluster mass, $\mathrm{R_h}$ and the larger $\mathrm{N_2/N_{tot}}$ ratio;
\item Migration of the cluster to larger galactocentric distances leads to an increase in cluster mass and $\mathrm{R_h}$. The $\mathrm{N_2/N_{tot}}$ ratio remains virtually constant after migration. 
\end{enumerate}
The reaction of the observational parameters to the changes in global and environmental parameters discussed in the above points is summarized in Table~\ref{t:Tab2}.

\begin{table}[]
\centering
\caption{Response of the observational global cluster parameters to changes in the models of the most important parameters describing the cluster environment and MSPs}
\begin{tabular}{cccc}
\hline\hline
Parameter              & M & $\mathrm{R_h}$ & $\mathrm{N_2/N_{tot}}$ \\ \hline
$\mathrm{R_g}$~$\uparrow$                  & = & $\uparrow$  & $\downarrow$       \\
$\mathrm{W_{o1}}$~$\uparrow$       & $\uparrow$ & $\uparrow$  & $\downarrow$    \\
$\mathrm{Q_1}$~$\uparrow$        & $\downarrow$ & $\downarrow$  & $\uparrow $  \\
migration~$\uparrow$      & $\uparrow$ & $\uparrow$  & =   \\  \hline 
\end{tabular}
\tablefoot{The global cluster parameters are: cluster mass, $R_h$ and $N_2/N_{tot}$). The parameters describing the cluster model and MSPs are: $\mathrm{R_g}$, $\mathrm{W_{o1}}$, $\mathrm{Q_1}$ and migration. ↑ means increase, ↓ means decrease, and = means unchanged.}
\label{t:Tab2}
\end{table}

Our goal is to determine the initial parameters of the models, including those related to the environment in which the clusters live, so that the observational parameters of the simulated clusters after 12 Gyr of evolution are within observational ranges and the correlation between the $\mathrm{N_2/N_{tot}}$ ratio and the cluster mass is recovered. Our analysis focuses only on in situ GCs for which the apparent circular orbits are less than \rm{7} kpc.

Globular clusters are formed in the very early stages of the galaxy's evolution, when the galaxy is a very hostile environment, with a highly changing tidal field and multiple close and strong interactions with nearby dwarf galaxies. The mass of the galaxy is rapidly being built up during this period. The work of \citet{Meng2022} suggests that the period of rapid changes in the tidal field of a galaxy ends after about 1 Gyr. After this time, GCs virtually stop migrating outside.
 
As many numerical simulations \citep[e.g.,][]{BanerjeeKroupa2013, Levequeetal2022} have shown, gas expulsion after the formation of \fg is associated with strong cluster expansion and excess of kinetic energy relative to potential energy ($\mathrm{Q} > 0.5$). Such a cluster will not exhibit a very high-density contrast (but rather small $\mathrm{W_{o1}}$, significantly less than 5 or 6), and for small galactocentric distances it should be TF. The gas expulsion effect is mainly related to the SFE and not to the position of the cluster in the galaxy. Thus, even in the case of a relatively large SFE \citep[e.g.,][]{Polaketal2024} we should expect some cluster expansion due to the removal of a significant mass of residual gas. So, clusters that arise closer to the center have smaller Rt than those that arise further away, and can also more easily become TF or even TF overfilling.

The joint action of cluster migration and initial models with $\mathrm{Q} > 0.5$ seems to work in the desired direction ($\mathrm{R_h}$ of a few parsecs and significant values of $\mathrm{N_2/N_{tot}}$), provided that the massive cluster is formed relatively close to the galactic center (about \rm{2-3} kpc or less), or the massive cluster experience stronger tidal field following its formation \citep[][]{Meng2022}. In order to obtain a model with a suitable range of observational parameters after a period of 12 Gyr of evolution, we should start with a TF cluster close to the center of the galaxy with a mass of at least $10^6 \mathrm{M_{\odot}}$ and a bit out of virial equilibrium. Unfortunately, the above initial conditions do not guarantee the observed correlation between GC mass and $\mathrm{N_2/N_{tot}}$ ratio. To try to reproduce such a correlation, one should assume that the availability of gas from which GCs can form decreases with distance from the galactic center. Thus, clusters forming farther from the center will have, on average, a lower mass of \fg and, more importantly, probably a significantly lower mass of gas that can again be accreted into the cluster and form \sg. As we know from the Section~\ref{s:Results}, the small value of $\mathrm{N_2}$ relative to $\mathrm{N_1}$ causes the increase in the $\mathrm{N_2/N_{tot}}$ ratio to be slow initially and only increase significantly after a period of 1 - 2 Gyr. Thus, if the cluster migrates during this time, its observed value of the $\mathrm{N_2/N_{tot}}$ ratio will be relatively low. For massive clusters formed from very massive gas clouds close to the galactic center, the availability of gas is very high, which will lead to the formation of clusters that initially have a relatively large $\mathrm{N_2/N_{tot}}$ ratio. For such clusters, the increase in this ratio to large values occurs quickly and then stabilizes. This means that their migration can stop the $\mathrm{N_2/N_{tot}}$ ratio at relatively large values. Dynamical friction also seems to be an important player in this process. We know that its effectiveness depends on the mass of the cluster and the distance from the center of the galaxy. Therefore, very massive clusters formed too close to the center or located after migration in not very eccentric orbits will be quickly “absorbed” by the nuclear star cluster. Those that have greater eccentricities and large semimajor axes will survive (as can be observed for certain MWGCs). Less massive clusters are likely to migrate to larger distances, and therefore dynamical friction will not be effective for them. Therefore, the mutual interaction of dynamical friction, migration, and availability of gas from which GCs can be formed may lead to the generation of the currently observed correlation between the mass of GCs and their $\mathrm{N_2/N_{tot}}$ ratio. 

This is a rather speculative and perhaps slightly naive refinement of the AGB scenario to form correlations observed for MW and galactic GCs. It requires several physical processes to operate in the right time order and in the right environment of the galaxy, which is just forming and assembling. Perhaps this answers the questions of why the correlation between $\mathrm{N_2/N_{tot}}$ and cluster mass is not reproduced by any proposed scenarios and why we do not see observational signatures of MSPs for the YMSCs we currently observe. It is simply that the current galactic environment is not suitable for the physical processes that enable the formation of MSPs with observed correlations.
 
Of course, the refinement provided above does not mean that this is the only way to explain the observed properties of MSPs in MWGC. In our view, they can provide a suitable environment for the formation of key MSP features. However, this is not the only possibility that can lead to the formation of MSP signatures. The scenarios discussed in the previous section may work simultaneously with the AGB scenario and lead to greater blurring of photometric and spectroscopic parameters of both \fg and \sg.  We would be surprised if the scenarios discussed above did not contribute to what is currently observed in MWGCs.

Finally, we note that according to \citet{Ishchenkoetal2024}, the orbits of MWGCs can undergo significant evolution over a Hubble time due to variations in the Galactic tidal field, resulting in either inward or outward migration.
This migration further blurs the initial parameters of the clusters and introduces a large scatter in their kinematic observational parameters. 

\section{Conclusions}
\label{s:Conclu}
In this work, we have extended the model of the evolution of GCs with MSPs presented in previous works \citet{Hypkietal2022, Hypkietal2025}. The model is anchored in AGB scenarios \citep[e.g.,][]{DErcoleetal2008, Conroy2011, Bekki2017, Caluraetal2019}. The possibility of a delay in time \sg formation has been added. Technically, in the \mocca code, the \sg formation process is defined by a number of parameters describing the properties of \fg, gas re-accretion, and \sg. The most important parameters are related to the migration of GCs in the galaxy, the mass of re-accreted gas, and the kinematic structure of \fg ($\mathrm{W_{o1}}$, $\mathrm{Q_1}$). All parameters are summarized in Table~\ref{t:Tab1}. 

Based on more than 100 simulations, we analyzed the influence of the most important parameters on the evolution of the cluster and its observational properties after 12 Gyr of evolution. The aim of the study is to determine such parameter ranges so that the cluster model after 12 Gyr of evolution best match the ranges of observational parameters of the MWGCs; in particular, the cluster mass, $\mathrm{R_h}$, and the $\mathrm{N_2/N_{tot}}$ ratio.  In addition, observational data collected from the following MWGCs catalogs were analyzed: \citet{Miloneetal2017} for $\mathrm{N_2/N_{tot}}$ ratios, \citet{Bajkova2021} for kinematic parameters, \citet{Chen2024} for origin types, and \citet{Baumgardtetal2019} for global parameters.  
The results of the work can be summarized as follows:
\begin{enumerate}
\item The time delay of gas re-accretion in the cluster relative to the \fg formation time causes the cluster to become nTF. This causes a significant slowdown in the growth of the $\mathrm{N_2/N_{tot}}$ ratio during evolution (see Figure \ref{f:Fig1});
\item Migration of GCs to larger galactocentric distances during the early time of galaxy formation causes a significant increase in $\mathrm{R_t}$, and thus the GC becomes nTF. Thus, the $\mathrm{N_2/N_{tot}}$ ratio practically stagnates (see Figure \ref{f:Fig2});
\item The ejection of gas after the formation of \fg is associated with a strong loss of cluster mass and potential energy. Thus, \fg becomes too hot and $\mathrm{Q} > 0.5$. The result of these processes is a very rapid increase in the $\mathrm{N_2/N_{tot}}$ ratio and a decrease in the mass of the cluster and $\mathrm{R_h}$. This makes it possible, despite the effects of migration and gas re-accretion, to obtain the cluster's observational parameters within the range of those observed for MWGCs (see Figure \ref{f:Fig7});
\item For a $\mathrm{N_2/N_1}$ ratio below 10\%, we observe the interesting behavior of the $\mathrm{N_2/N_{tot}}$ ratio evolution. In the initial period up to about 1-2 Gyr, it remains practically constant in order to increase rapidly later and reach the values observed in MWGCs (see Figure \ref{f:Fig3}). The properties of these models hopefully allow us to better understand why YMSCs do not exhibit MSP signatures; 
\item The GC should form near the center of its host galaxy and be very massive (at least about $10^6 M_{\odot}$), so after removing the residual \fg gas, it should be TF or close to TF. Close proximity to the galactic center means smaller $\mathrm{R_t}$ and much easier filling of the Roche lobe after the ejection of residual gas. Only GCs that are initially TF can significantly increase the $\mathrm{N_2/N_{tot}}$ ratio during evolution.
\end{enumerate}

Based on the above and the results of cosmological zoom-in simulations, a speculative refinement of the AGB scenario for the formation and evolution of GCs with MSP was presented to obtain an observational correlation between cluster mass and the $\mathrm{N_2/N_{tot}}$ ratio. To achieve this, one must combine the influence of the strongly changing MW environment in which the cluster evolves with its internal evolution and the effects related to the ejection of the residual gas after \fg formation. As is discussed in Section \ref{s:Specul}, all those physical processes have to operate in the right time
order and in the right environment of the galaxy that is just forming and assembling. In order to confirm this speculative refinement, more sophisticated modeling of the evolving tidal field in which the GCs form and evolve is required. This will require high-resolution cosmological zoom-in simulations \citep[e.g.,][]{Lietal2017, Renaud2020,Caluraetal2024}.

It is very important to emphasize that models of GCs with MSPs based on the AGB scenario require completely different initial models than models of clusters with a single population. Instead of GCs being highly concentrated and lying deep inside the Roche lobe, models that fill the Roche lobe are required. This carries strong constraints on where in the galaxy GCs are formed.

Work is currently underway to further expand the \mocca code to include the ability to take into account GC motion in the galaxy's real, time-varying potential. For this purpose, based on cooperation with Peter Berczik's team, the machinery described in the work of \citet{Ishchenkoetal2024} will be used for integrating GC orbits in a MW-type galaxy, a galaxy that is currently forming and building its mass. The time-varying GC orbit (semimajor axis and eccentricity) will be used to determine the effective tidal radius and mass loss of the GC. It is hoped that this will track GC migration and its impact on the $\mathrm{N_2/N_{tot}}$ ratio and other observational GC parameters.


\begin{acknowledgements} 
We are grateful to an anonymous reviewer for his/her insightful comments, which helped us improve the article.
GM, AH, AA, GW, and LH were supported by the Polish National Science Center (NCN) through the grant 2021/41/B/ST9/01191.
AA acknowledges support for this paper from project No. 2021/43/P/ST9/03167 co-funded by the Polish National Science Center (NCN) and the European Union Framework Programme for Research and Innovation Horizon 2020 under the Marie Skłodowska-Curie grant agreement No. 945339. For the purpose of Open Access, the authors have applied for a CC-BY public copyright license to any Author Accepted Manuscript (AAM) version arising from this submission.
\end{acknowledgements} 
 

\section*{Data Availability}
The simulation data underlying the results of this study are publicly available at Zenodo: \href{https://doi.org/10.5281/zenodo.15379060}{https://doi.org/10.5281/zenodo.15379060}. The dataset includes MOCCA \texttt{system.dat} files for models discussed in the paper, organized by initial cluster properties and simulation parameters. 

\bibliographystyle{aa}
\bibliography{ref.bib}

\end{document}